\documentclass[12pt]{amsart}

\usepackage[english]{babel}
\usepackage[
    letterpaper,
    top=3cm,
    bottom=3cm,
    left=2cm,
    right=2cm,
    marginparwidth=1.5cm
]{geometry}

\usepackage{cite}
\usepackage{amsmath}
\usepackage{amssymb}
\usepackage{amsfonts}
\usepackage{graphicx}
\usepackage{float}
\usepackage{subfig}
\usepackage[T1]{fontenc}
\usepackage[colorlinks=true, allcolors=blue]{hyperref}

\newcommand*{\dd}{\mathrm{d}}
\newcommand*{\ii}{\mathrm{i}}

\DeclareMathOperator{\sn}{sn}
\DeclareMathOperator{\ns}{ns}

\begin{document}
\title{The double Schwarzschild solution in bispherical coordinates}
\author[C. Klein]{Christian Klein}
\address[C.~Klein]
{Université Bourgogne Europe, CNRS, IMB UMR 5584, 21000 Dijon, 
France,
Institut Universitaire de France}
\email{christian.klein@u-bourgogne.fr}

\author{El Mehdi Zejly}
\address[M.~Zejly]
{Université Bourgogne Europe, CNRS, IMB UMR 5584, 21000 Dijon, France}
\email{El-Mehdi.Zejly@u-bourgogne.fr}

\begin{abstract}
	The double Schwarzschild solution in the equal mass case is 
	studied in bispherical coordinates. An explicit conformal transformation 
	from cylindrical Weyl coordinates to bispherical coordinates is 
	given in terms of elliptic functions. A multi-domain spectral method for spacetimes in 
	bispherical coordinates is presented to numerically reconstruct 
	this solution.  
\end{abstract}

\date{\today}


\thanks{This work was supported by the ANR project 
ANR-17-EURE-0002 EIPHI and by the ANR project 
ISAAC-ANR-23-CE40-0015-01. We thank J.~Frauendiener for helpful 
discussions and hints. }

\maketitle
\section{Introduction}
Binary black holes systems are among the strongest sources for 
gravitational radiation in our universe, and consequently, the first 
gravitational waves to be recorded in \cite{abbott2016} originate 
from such a system. The Einstein equations for binary black hole 
systems are complicated and can only be solved numerically for the 
time being. In an earlier 
stage of the evolution of such a system, a quasi-stationary phase is 
expected where the change of radius due to emitted radiation is small over 
one  complete turn. Detweiler 
\cite{detweiler1989,blackburn1992,detweiler1994} suggested to 
approximate this stage by a spacetime with a \emph{helical symmetry} 
where the outgoing radiation is exactly compensated by incoming 
gravitational waves. Such an approach  had
been previously applied  to binary charges of opposite sign
in Maxwell theory by Sch\"onberg \cite{schonberg1946} and Schild 
\cite{schild1963}.

In a numerical context, spacetimes with a helical Killing vector have 
been studied in 
\cite{friedman2002,gourgoulhon2002a,grandclement2002,andrade2003,caudill2006,uryu2009,friedman2006,bishop2005,uryu2010,yoshida2006,beetle2006,bromley2005,beetle2007,lau2007,lau2012,bonazzola2007} and 
references therein. A formulation of the Einstein equations in vacuum 
in the presence of a helical Killing vector was given in 
\cite{klein2004} by applying Ehler's projection formalism 
\cite{Ehlers1957}, see also \cite{Geroch1971}. This led to a 
formulation of the Einstein equations in the form of an Ernst 
equation \cite{Ernst1968}, see also the discussion in 
\cite{KleinRichter2005}. In this formalism, binary black hole spacetimes with a helical 
Killing vector are potentially singular at the horizons of the black 
holes that are Killing horizons and the light cylinder, where 
observers stationary with respect to the horizons rotate with the 
velocity of light. Null infinity cannot be regular in these 
spacetimes, see \cite{gibbons1984,ashtekar1978} since there will be 
incoming radiation to compensate the emitted radiation from the 
binary system. 
In the Ernst formalism the zeros of the Killing 
vector  lead to Fuchsian singularities of the resulting equations 
which where studied in a formal expansion in \cite{klein2004}. 

In addition to the potential application of helical Killing vectors 
in an early stage of a binary system, there is also a mathematical 
interest in such systems since they are of mixed type (elliptic in the interior of the light cylinder and hyperbolic in the
exterior). To construct such spacetimes 
numerically, it will be necessary to solve a 3D system of nonlinear equations 
with Fuchsian singularities at the boundary and non regular infinity. 
As a first step towards this problem, we studied numerically in 
\cite{Bai2016} the case of a single Kerr black hole in a rotating 
frame. 
Since the angular velocity of the frame was chosen to equal the formal 
angular velocity of the horizon, this corresponded to a situation 
where the horizon is a Killing horizon, and where there is a second 
surface of cylindrical topology, the light cylinder, where the Killing 
vector changes its causal nature. Since the solution is explicitly known, this 
provided an interesting test of the numerical algorithms.

In the present paper we want to consider an explicitly known test 
case with two Killing horizons. To this end we consider the well 
known static double Schwarzschild solution \cite{KramerNeugebauer1980}, see 
also \cite{Kramer1980}, where the black holes are separated by a 
conic singularity on the axis, a \emph{Weyl strut}. This spacetime 
was studied via ray tracing in \cite{deLeon2025}. The solution is 
given in cylindrical Weyl coordinates where the horizons form 
intervals on the symmetry axis. Since this is numerically not a 
convenient setting, we establish first an exact conformal 
transformation to \emph{bispherical coordinates} where the horizons 
of spherical topology are on constant coordinate surfaces. In these coordinates 
infinity is compactified to a point in the 
computational domain. An explicit form of the double Schwarzschild solution  in 
bispherical coordinates is given for the first time 
in terms of Jacobi elliptic functions. We 
numerically reproduce this solution  with an approach similar to  
Grandcl\'ement's \emph{Kadath} \cite{Grandclement2010}, a 
multi-domain spectral approach. It is shown that the regular part of 
the exactly known metric can be reproduced numerically to machine 
precision with this approach. 

The paper is organised as follows: in section 2 we review stationary 
axisymmetric spacetimes in the Ernst formalism and the double 
Schwarzschild solution in Weyl coordinates. In section 3 we present a 
short overview on bispherical coordinates. In section 4 we construct 
the conformal transformation from Weyl coordinates to bispherical 
ones for the double Schwarzschild solution. The numerical methods to 
be used are outlined in section 5. In section 6 we solve the Einstein 
equations in bispherical coordinates for the double Schwarzschild 
solution. In section 7, we add some concluding remarks. Details on 
the solution in terms of elliptic functions are presented in an 
appendix. 

\section{Double Schwarzschild solution in Weyl coordinates}
\label{sec:double-schw-solut}
In this section, we collect a few facts on static axisymmetric 
vacuum spacetimes and the  double 
Schwarzschild solution. 

In the static axisymmetric case in vacuum, the metric can be written 
in Weyl-Lewis-Papapetrou form, see \cite{Kramer1980},
\begin{equation} 
	\dd s^{2}=-f 
	\dd t^{2}+\frac{1}{f}\left(e^{2k}(\dd\rho^{2}+\dd z^{2})+\rho^{2} 
	\dd \phi^{2}\right);
	\label{metric}
\end{equation}
here the cylindrical Weyl coordinates are such that $\rho$ measures 
the distance to the symmetry axis parameterized by $z$, and 
$\partial_t$ and $\partial_\phi$ correspond to the static and the 
axisymmetric Killing vector respectively. The metric functions $f$ 
and $k$ depend on $\rho$ and $z$, but not on $\phi$ and $t$. The 
potential $\ln f$ satisfies the axisymmetric Laplace (or Euler-Darboux) 
equation, 
\begin{equation}
	(\ln f)_{\rho\rho}+\frac{1}{\rho}(\ln f)_{\rho}+(\ln f)_{zz}=0
	\label{ED}.
\end{equation}
Writing the metric function $g_{\phi\phi}$ in the form 
$\mathcal{W}^{2}=fg_{\phi\phi}$, one finds that, from Einstein's equations, $W$ satisfies the equation 
\begin{equation}
	\mathcal{W}_{\rho\rho}+\mathcal{W}_{zz}=0
	\label{Weq},
\end{equation}
i.e., the 2D Laplace equation. An obvious solution vanishing on the axis is $\mathcal{W}=\rho$. 
It is convenient to introduce complex notation, $\xi:=z-\ii\rho$. In 
this case the metric function $k$ is given for known $f$ and $\mathcal{W}$ in terms of quadratures,
\begin{equation}
  k_{\xi}=\frac{(\xi-\bar{\xi})}{4}
  (\ln f)_{\xi}^{2}\;.
  \label{kxi}
\end{equation}
In other words, the Einstein equations in vacuum are equivalent to 
$\mathcal{W}$ being a harmonic function, $\ln f$ satisfying the Euler-Darboux 
equation and $k$ being given in terms of quadratures of these 
functions. The complex 
notation introduced above will facilitate the finding of a conformal 
transformation to bispherical coordinates below.

It is well known that the static axisymmetric Einstein equations 
in vacuum are  completely integrable, see for instance 
\cite{Ernst1968,Kramer1980,KleinRichter2005} for references. 
Mathematically so-called multi-black 
hole  solutions are multi-solitons. In the static case, they 
correspond to exact solutions having disconnected horizons on the 
symmetry axis where $f$ vanishes on intervals.  Here we concentrate 
on the equal mass case (for which the solution has an equatorial 
symmetry), where we follow the presentation in \cite{MankoRuiz2019}.

\begin{figure}[H]
  \includegraphics[width=0.49\textwidth]{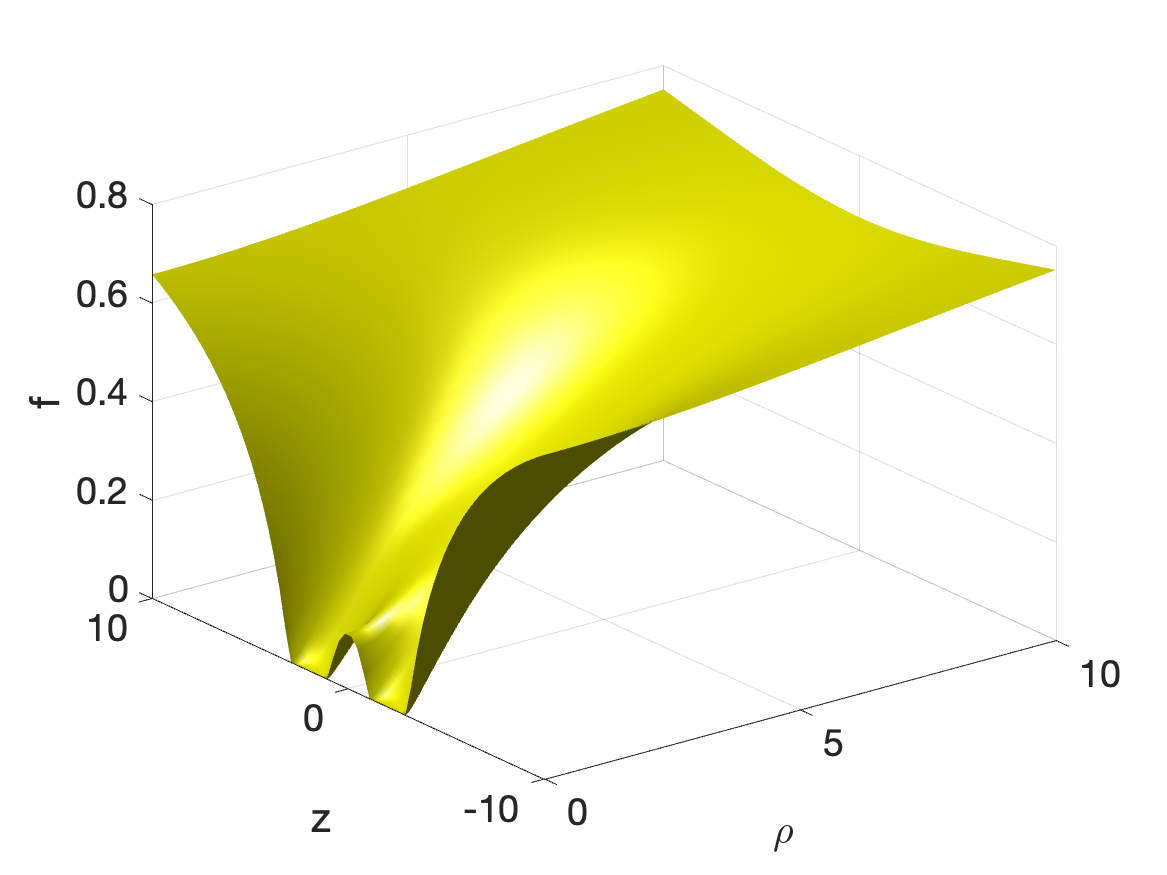}
  \includegraphics[width=0.49\textwidth]{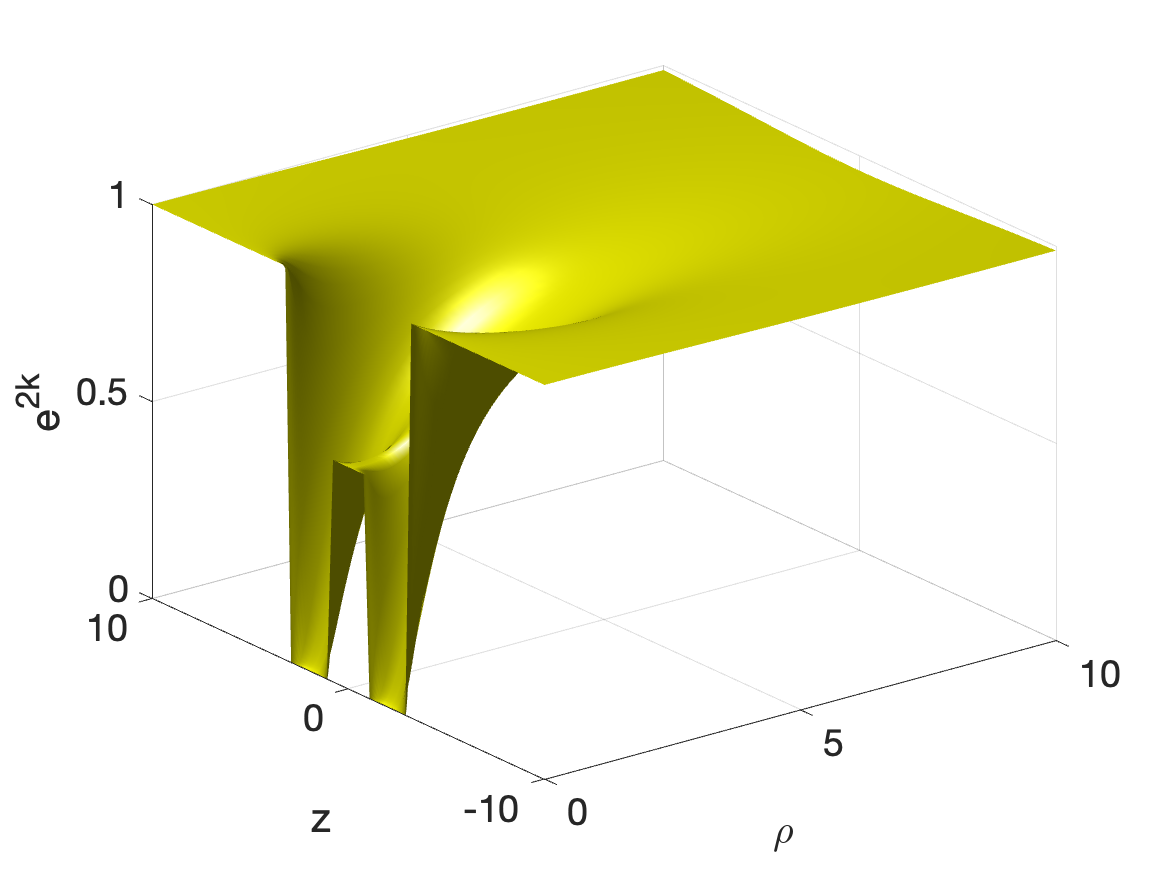}
 \caption{The metric functions (\ref{fk}) for the double 
 Schwarzschild solution for $m_{1}=m_{2}=1$ and $R_{0}=4$, on the left 
 $f$, on the right $e^{2k}$.}
 \label{DSm1}
\end{figure}

The solution is parameterized by three 
parameters $m_{1}$, $m_{2}$ corresponding to the Komar masses of the 
two black holes (put $M=m_{1}+m_{2}$), and a distance $R_{0}>m_{1}+m_{2}$.
We have
\begin{equation}
	R_{\pm}=\mp\sqrt{\rho^2+(z+R_{0}/2\pm m_{2})^2},\quad 
	r_{\pm}=\mp\sqrt{\rho^2+(z-R_{0}/2\pm m_{1})^2},
	\label{rR}
\end{equation}
as well as 
\begin{equation}
  \begin{aligned}
	A & 
	=(R_{0}^2-M^2)(R_{+}-R_{-})(r_{+}-r_{-})-4m_{1}m_{2}(R_{+}-r_{-})(R_{-}-r_{+}),
	\\
	B & = 2m_{1}(R_{0}^2-m_{1}^2 + m_{2}^2)(R_{-}-R_{+}) +  
	2m_{2}(R_{0}^2-m_{2}^2 + m_{1}^2)(r_{-}-r_{+}) \\
        & \hspace*{3em}+
    4R_{0}m_{1}m_{2}(R_{+}+R_{-}-r_{+}-r_{-}).
	\label{AB}
      \end{aligned}
\end{equation}
Then the metric functions for the double Schwarzschild solution can 
be written as 
\begin{align}
	f & 
	=\frac{A-B}{A+B},
	\nonumber\\
	e^{2k} & = 
	\frac{(A^2-B^2)}{16R_{+}R_{-}r_{+}r_{-}(R_{0}^2-(m_{1}-m_{2})^2)^2}.
	\label{fk}
\end{align}
We show the metric functions for 
the example $m_{1}=m_{2}=1$ and $R_{0}=4$ in Fig.~\ref{DSm1}. 

There are two horizons 
located  on the symmetry axis between $R_{0}/2\pm m_{1}$ and 
$-R_{0}/2\pm m_{2}$.  Both functions, $f$ and $e^{2k}$, vanish there. The 
solution is asymptotically flat. On the regular part of 
the axis, here for $|z|>3$, the metric function $k$ vanishes. However,
this is not the case between the two horizons, for $|z|<1$ in the 
example, where, $k$ is constant, but  not equal to zero. 
This corresponds to a conical singularity in the space-time, a \emph{Weyl 
strut}, which keeps the situation static despite the two black holes 
attracting each other.

\section{Bispherical coordinates}
In this section we collect some basic facts on bispherical 
coordinates.
Since the double Schwarzschild metric contains two distinct horizons with spherical topology, it is natural to introduce bispherical coordinates for the description of this spacetime. Starting from Cartesian coordinates $(x,y,z)$, bispherical coordinates are defined as follows. Let
\[
F_1 := (0,0,a), \qquad F_2 := (0,0,-a),
\]
where $a>0$ is a parameter; these points are called the focal points of the bispherical coordinate system. For any point $P\in\mathbb{R}^3$, define
\[
    \eta := \ln\left(\frac{d(P,F_1)}{d(P,F_2)}\right)\in\mathbb{R},
    \qquad
    \theta := \measuredangle F_1PF_2 \in [0,\pi].
\]
\begin{figure}[H]
\includegraphics[width=0.8\textwidth]{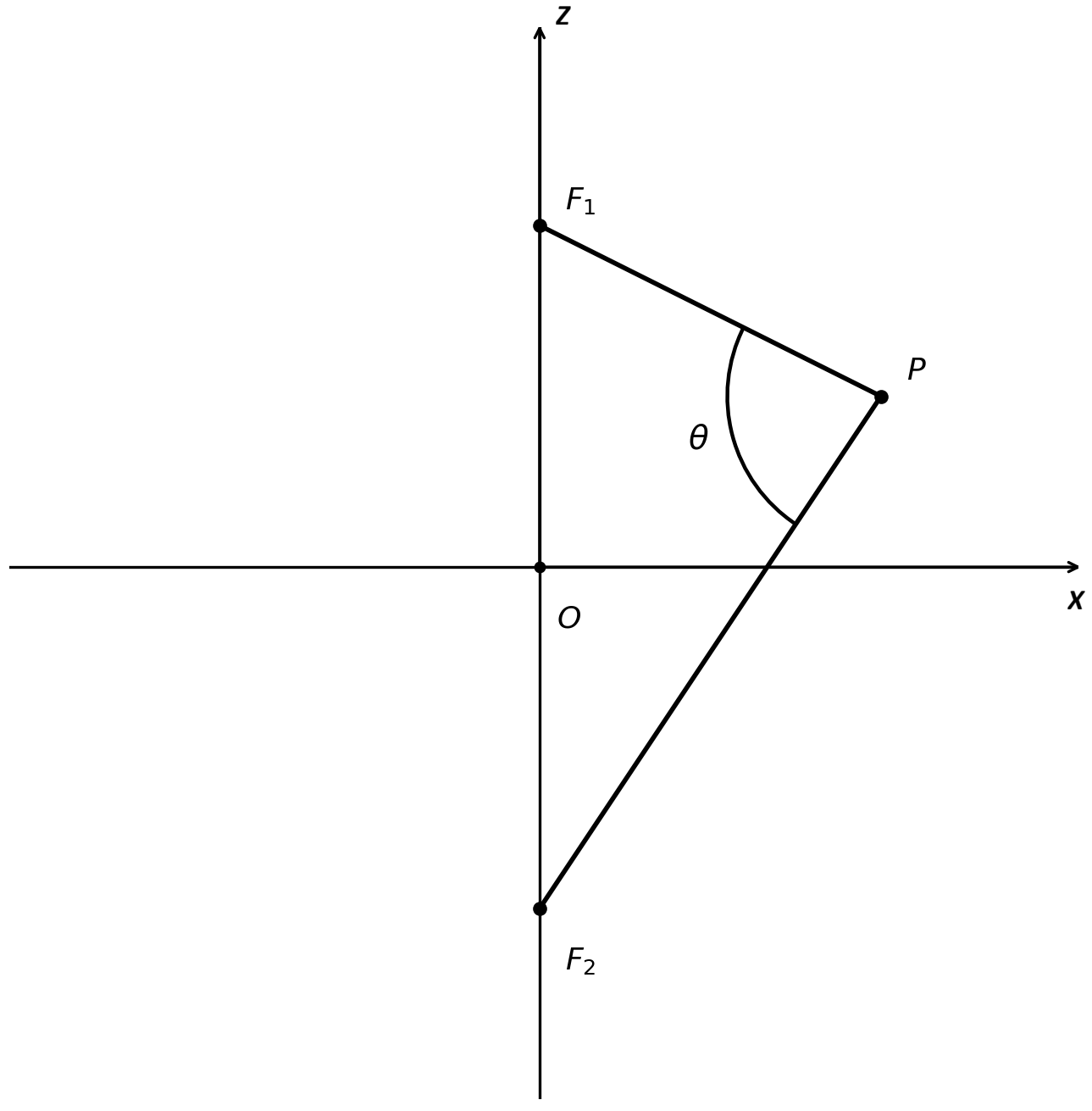}
 \caption{Illustration of the angle $\theta$ at point $P$ in the $x$-$z$ plane.}
 \label{bipolar}
\end{figure}
Here, $d(P,F_1)$ and $d(P,F_2)$ denote the Euclidean distances from $P$ to $F_1$ and $F_2$, respectively. If $\psi\in[0,2\pi)$ denotes the azimuthal angle around the $z$-axis, then $(\eta,\theta,\psi)$ are called the bispherical coordinates of $P$. These coordinates are described in more detail in \cite{fth_pm_des_61}. The surfaces $\{\eta=\mathrm{const.}\}$ are nested spheres with center $(0,0,a\coth\eta)$ and radius
\[
\frac{a}{|\sinh\eta|},
\]
except in the case $\eta=0$, which corresponds to the plane $\{z=0\}$.

\begin{figure}[H]
\includegraphics[width=0.45\textwidth]{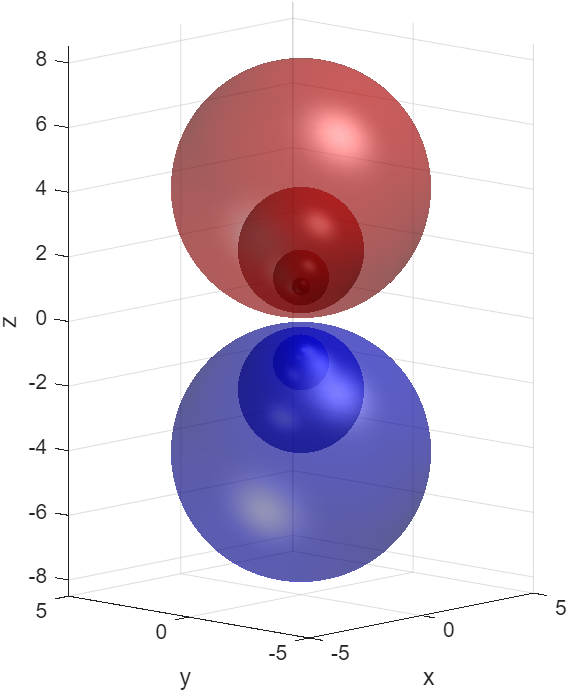}
 \caption{The surfaces $\{\eta=\text{const.}\}$ for $\eta=0.25$, $0.5$, $1$, $2$, and $3$ (shown in red), together with the corresponding symmetric surfaces for negative values of $\eta$ (shown in blue), with $a=1$.}
 \label{EtaConst3D}
\end{figure}

\begin{figure}[H]
\includegraphics[width=0.65\textwidth]{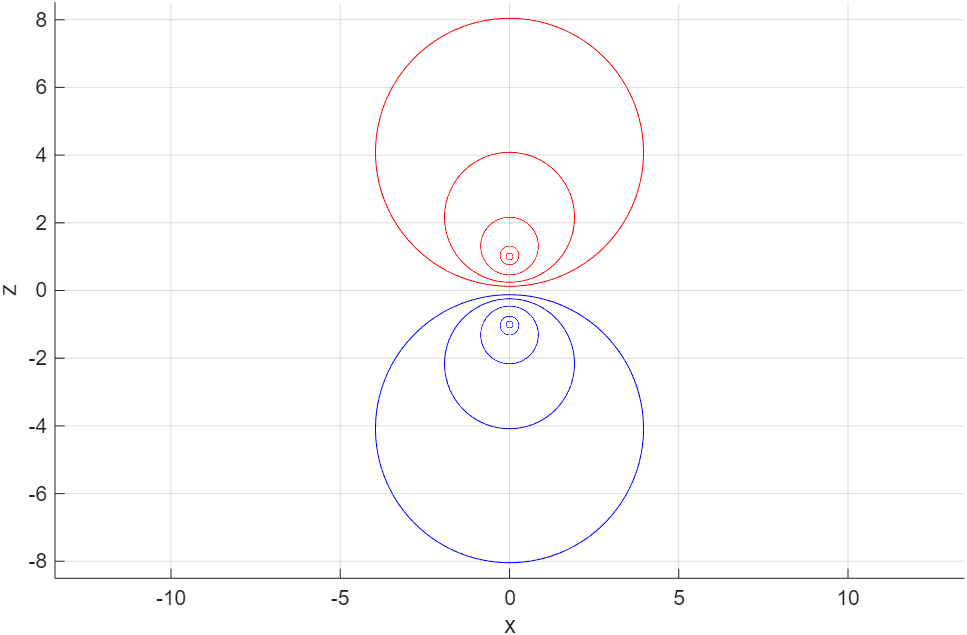}
 \caption{The sections of the surfaces $\{\eta=\text{const.}\}$ in the plane spanned by the $x$- and $z$-axes, for the same values as in Fig.~\ref{EtaConst3D}.}
 \label{EtaConst2D}
\end{figure}

In these coordinates, the Cartesian coordinates are given by
\begin{equation}
    x = \frac{a\sin\theta\cos\psi}{Q},
    \qquad
    y = \frac{a\sin\theta\sin\psi}{Q},
    \qquad
    z = \frac{a\sinh\eta}{Q},
    \label{cart_bis}
\end{equation}
where
\begin{equation}
    Q := \cosh\eta - \cos\theta.
    \label{Q}
\end{equation}

In the axisymmetric case, by setting the radial coordinate $\rho^2 := x^2 + y^2$, one has
\begin{equation}
	\rho= \frac{a\sin\theta}{Q},
    \qquad
    z = \frac{a\sinh\eta}{Q},
	\label{bisphericalrho}
\end{equation}
and 
\begin{equation}
	\rho^{2}+z^{2}=a^2\frac{\cosh\eta+\sin\theta}{Q}.
	\label{inf}
\end{equation}
This implies that infinity is reached in these coordinates in the 
limit $\eta\to0$ and $\theta\to0$. Thus bispherical coordinates lead 
in a natural way to a compactification of $\mathbb{R}^{2}$. 

\section{Metric in bispherical coordinates}

For the equal-mass case, Weyl and bispherical coordinates are related by the conformal map constructed in Appendix~A:
\[
\rho+\ii z = w(u), \qquad u:=\eta+\ii\theta,
\]
with
\begin{equation}
    w(u) = \ii\left(\frac{R_0}{2}+m\right)\ns\left(\frac{K}{\eta_0}u, \mu\right),
    \label{w(u)}
\end{equation}
where
\[
\mu = \left(\frac{R_0-2m}{R_0+2m}\right)^2,
\qquad
\eta_0 = \frac{\pi K(\mu)}{K'(\mu)},
\]
and where $\mathrm{ns}(x)=1/\mathrm{sn}(x)$, with $\mathrm{sn}(x)$ 
being the standard Jacobi elliptic function (for a detailed 
definition see appendix A or \cite{Lawden}).
The horizons are located at $\eta=\pm\eta_0$, and spatial infinity corresponds to the pole $u=0$.
\begin{figure}[H]
    \centering
    \subfloat{{\includegraphics[scale=0.5]{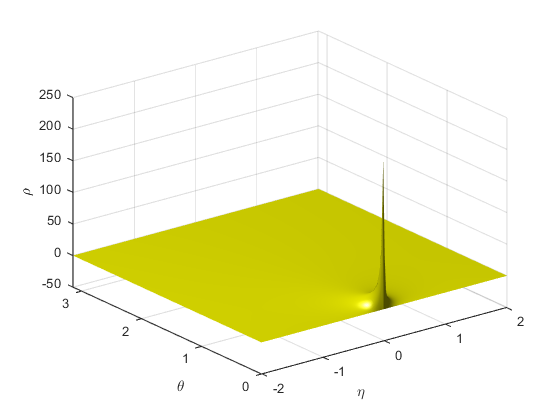}}}
    \qquad\qquad\qquad
    \subfloat{{\includegraphics[scale=0.5]{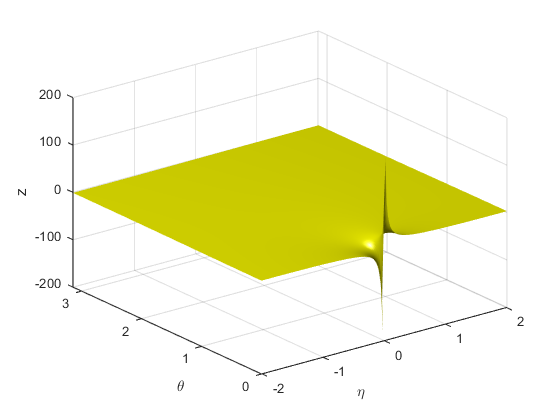}}}
    \caption{Weyl coordinates as functions of the bispherical coordinates: on the left, $\rho$, and on the right, $z$. The pole at $u = 0$ is clearly visible.}
\end{figure}

In order to better observe how the Weyl coordinates vary with respect to the bispherical coordinates, it is convenient to normalize them by the factor $Q$ defined in \eqref{Q}, since it vanishes at the pole with the same order, as shown in Appendix~C.

\begin{figure}[H]
    \centering
    \subfloat{{\includegraphics[scale=0.5]{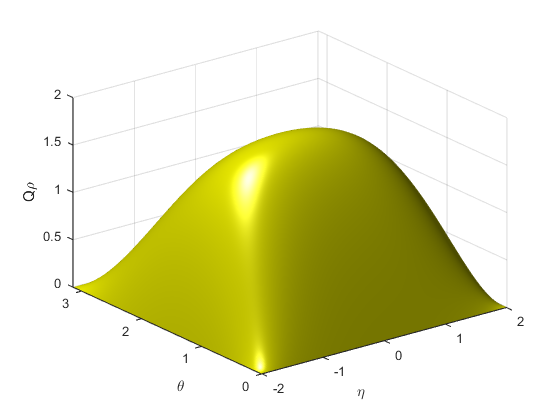}}}
    \qquad\qquad\qquad
    \subfloat{{\includegraphics[scale=0.5]{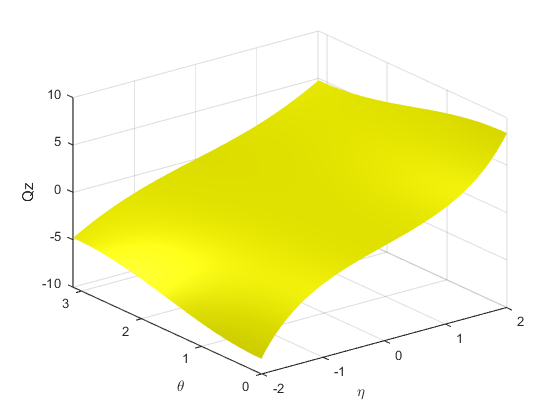}}}
    \caption{Normalized Weyl coordinates as functions of the bispherical coordinates: on the left, $Q\rho$, and on the right, $Qz$.}
    \label{norm_weyl}
\end{figure}

Denoting by
\begin{equation}
    h_{ab}\dd x^a\dd x^b := e^{2k}(\dd\rho^2 + \dd z^2) + \rho^2\dd\phi^2
    \label{space_metric}
\end{equation}
the spatial part of the metric, up to the conformal factor $f^{-1}$, in Weyl coordinates, one finds that the metric in bispherical coordinates can be written as
\[
    h_{\eta\eta} = h_{\theta\theta} = |w'(\eta+i\theta)|^2e^{2k},
    \qquad
    h_{\phi\phi} = \rho^2(\eta,\theta).
\]
From the expression for $w(u)$ in \eqref{w(u)}, one finds
\begin{equation}
    |w'(u)|^2 = \alpha^2\left(\frac{R_0}{2}+m\right)^2\left|\frac{\left(1-\sn^2\left(\alpha u,\mu\right)\right)\left(1-\mu\sn^2\left(\alpha u,\mu\right)\right)}{\sn^4\left(\alpha u,\mu\right)}\right|,
    \label{dw2}
\end{equation}
where $\alpha := \dfrac{K}{\eta_0}.$

\begin{figure}[H]
    \centering
    \subfloat{{\includegraphics[scale=0.5]{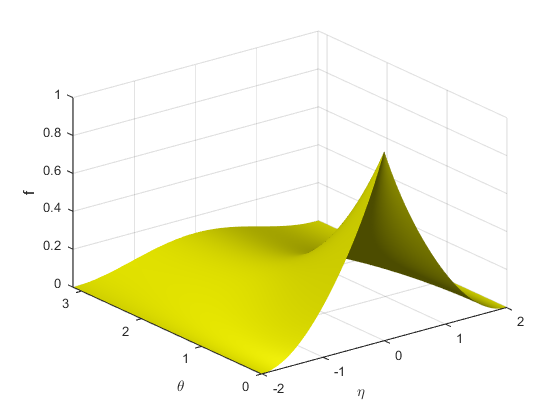} }}
    \qquad\qquad\qquad
    \subfloat{{\includegraphics[scale=0.5]{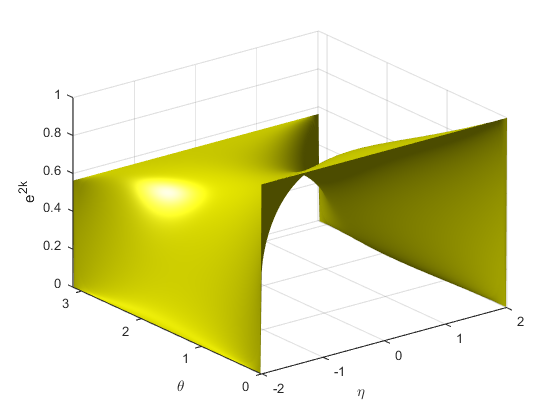} }}
   \caption{Weyl potentials as functions of the bispherical coordinates: on the left, $f$, and on the right, $e^{2k}$.}
   \label{fkbispfig}
\end{figure}

We show the metric potentials $f$ and $e^{2k}$ in bispherical 
coordinates in Fig.~\ref{fkbispfig}. They vanish on the horizons 
$\eta = \pm\eta_0$. The function $f$ has a visible cusp for 
$\eta=\theta=0$, the point corresponding to spatial infinity.  
The quantity $e^{2k}$ is discontinuous at the corners of the domain $\mathcal{D}_b$. It is also less than $1$ on the axis between the black holes at $\theta = \pi$, which corresponds to the Weyl strut.

Note that the boundary of the domains in Fig.~\ref{fkbispfig} 
corresponds in Weyl coordinates to the axis: the two horizons are now 
given by $\eta=\pm\eta_{0}$, the regular part of the axis including 
infinity to $\theta=0$, and the part of the axis between the horizons 
where the Weyl strut is located at $\theta=\pi$. 

\begin{figure}[H]
    \centering
    \subfloat{{\includegraphics[scale=0.45]{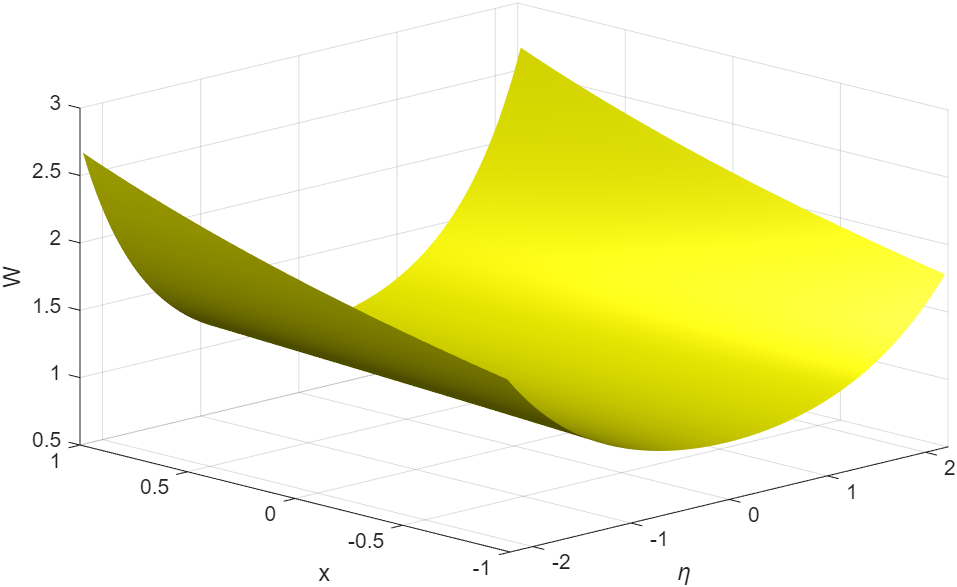}}}
    \qquad\qquad\qquad
    \subfloat{{\includegraphics[scale=0.45]{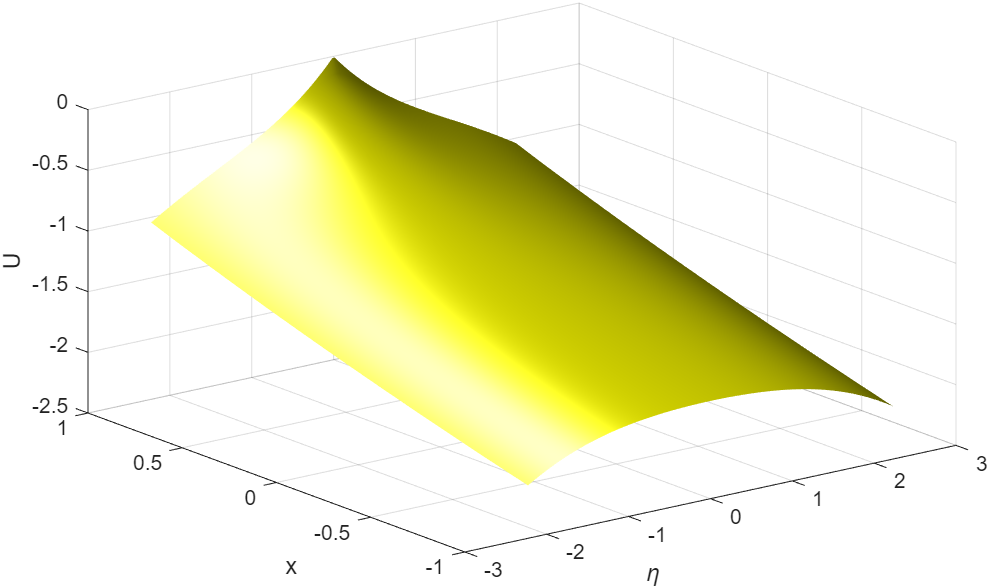}}}
    \caption{The functions $W$ (left) and $U$ (right) as defined in \eqref{rho} and \eqref{ernst2} respectively.}
    \label{W_and_U}
\end{figure}

\section{Numerical approach in a single domain}
In this section we introduce the numerical approach to be applied in 
the following, a Chebyshev collocation method as in \cite{Bai2016} and \cite{Grandclement2010}. With this approach in a 
single domain, we reconstruct the metric function $h_{\phi\phi}$. 

\subsection{Chebyshev collocation method}

The basic idea of a Chebyshev collocation method is to map an 
interval $[a,b]$ for some variable $x$ via $x=b(1+l)/2+a(1-l)/2$ to 
the interval $[-1,1]$ where $l\in[-1,1]$. For $l$ we introduce 
the usual Chebyshev collocation points $l_{n}=\cos(n\pi/N)$, 
$n=0,1,\ldots,N$, $N\in \mathbb{N}$. Standard Lagrange interpolation 
of a function $u(l)$ on these collocation points allows for an approximation of the derivative of the function $u$ via the derivative of 
the interpolation polynomial, which is equivalent to the action of 
Chebyshev differentiation matrices on the vector with components 
$u(l_{0}),\ldots,u(l_{N})$, see \cite{Trefethen2000}. Note that the 
conditioning of Chebyshev differentiation matrices is of the order of 
$\mathcal{O}(N^{2})$. 

This approach is equivalent to the expansion of the function $u$ in 
terms of Chebyshev polynomials $T_{n}(l)=\cos(n\arccos(l))$, 
$n=0,1,\ldots$, 
\begin{equation}
	u(l)\approx \sum_{n=0}^{N}a_{n}T_{n}(l)
	\label{Tn}.
\end{equation}
In a collocation approach, the \emph{spectral coefficients} $a_{n}$, 
$n=0,\ldots,N$ are obtained by 
imposing relation (\ref{Tn}) as an equality on the collocation points,
\begin{equation}
	u(l_{k})= 
	\sum_{n=0}^{N}a_{n}T_{n}(l_{k})=\sum_{n=0}^{N}a_{n}\cos(nk\pi/N),\quad k=0,1,\ldots,N
	\label{an}.
\end{equation}
The spectral coefficients can be 
computed with a \emph{Fast Cosine Transform} (FCT) that is related to 
the Fast Fourier Transform. It is known, see for instance 
\cite{Trefethen2000}, that they decrease 
exponentially with $n$ for analytic functions $u(l)$ which is called 
\emph{spectral convergence}. Thus the 
numerical error is indicated by the coefficients with the highest 
index. We always aim at a numerical resolution of the order of 
machine precision which is in Matlab roughly $10^{-16}$ (due to the 
mentioned conditioning of the differentiation matrices, achievable 
accuracy is typically limited to $10^{-12}$). 

Boundary or matching conditions are implemented with Lanczos' 
$\tau$-method, see \cite{Trefethen2000}. This means that the boundary 
conditions are discretised in the same way as the to be solved 
differential equation, and that the line(s) in the differentiation 
matrices corresponding to the  boundary are replaced by the 
condition. 

In two spatial dimensions, say $x\in[-1,1]$ and $y\in[-1,1]$, one simply takes 
a tensor grid: 
$u(x,y)\approx\sum_{n=0}^{N_{x}}\sum_{m=0}^{N_{y}}a_{nm}T_{n}(x)T_{m}(y)$, and similarly for the differentiation matrices. 

\subsection{The potential $h_{\phi\phi}$}
If a conformal transformation of the standard Weyl cylindrical 
coordinates to bispherical coordinates is considered, the form of the 
Einstein equations does not change in the quotient space approach. 
Thus we have that $\rho$ is again a harmonic function. For the 
numerical treatment, we take explicit care of the singularities of 
this function which vanishes at the horizons $\eta=\pm\eta_{0}$ and 
the axis $\theta=0$. Thus we make the following ansatz for $\rho$,
\begin{equation}
	\rho = (\eta_{0}^{2}-\eta^{2})\frac{\sin\theta}{Q}W,
	\label{rho}
\end{equation}
motivated by the form of $\rho$ in bispherical coordinates in flat 
space.

With this ansatz we get for the 2D Laplace equation
\begin{equation}
	\begin{split}
	&W_{\theta\theta}+W_{\eta\eta}+2\left(\cot\theta-\frac{\sin\theta}{Q}\right)W_{\theta}
	-2\left(\frac{2\eta}{\eta_{0}^{2}-\eta^{2}}+\frac{\sinh\eta}{Q}\right)W_{\eta}\\
		&+\frac{W}{\eta_{0}^{2}-\eta^{2}}\left(\frac{4\eta\sinh\eta}{Q}-2\right) = 0
	\end{split}
		\label{W1}.
\end{equation}
It is convenient to introduce the coordinate $x=\cos\theta$ which 
leads for (\ref{W1}) to a less singular equation, 
\begin{equation}
	\begin{split}
	&W_{\eta\eta}+(1-x^{2})W_{xx}-3xW_{x}+2\frac{1-x^{2}}{Q}W_{x}
	-2\left(\frac{2\eta}{\eta_{0}^{2}-\eta^{2}}+\frac{\sinh\eta}{Q}\right)W_{\eta}\\
		&+\frac{W}{\eta_{0}^{2}-\eta^{2}}\left(\frac{4\eta\sinh\eta}{Q}-2\right) = 0
	\end{split}
		\label{W2}.
\end{equation}
This is a singular linear equation, the singularities being $x=\pm1$, 
$\eta=\pm\eta_{0}$ and $Q=0$. The equation is homogeneous in $W$. 
This means that a unique solution is 
expected as for instance in the case of the hypergeometric equation 
if the value of the solution is fixed at some of the singularities as 
in \cite{Crespo}.

\subsection{Numerical solution for the function $W$}
\begin{figure}[h!]
  \includegraphics[width=0.49\textwidth]{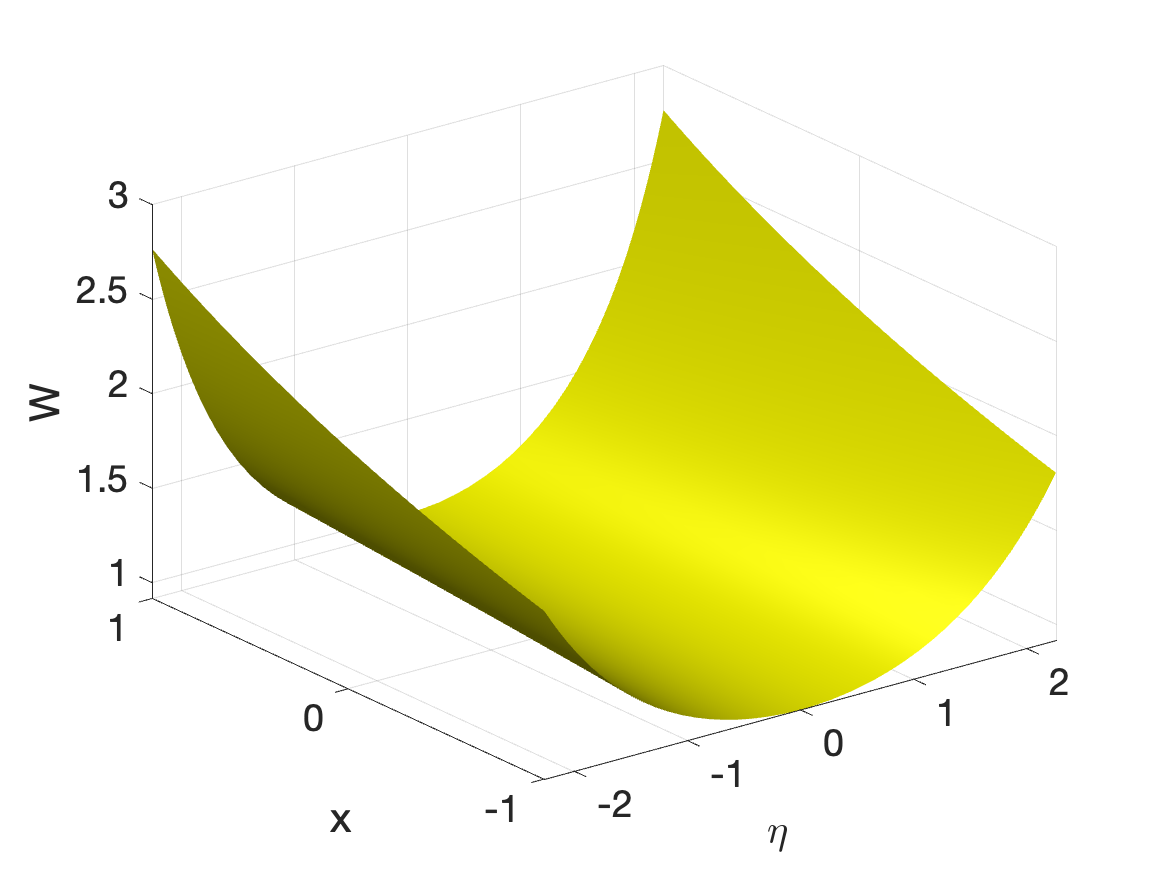}
  \includegraphics[width=0.49\textwidth]{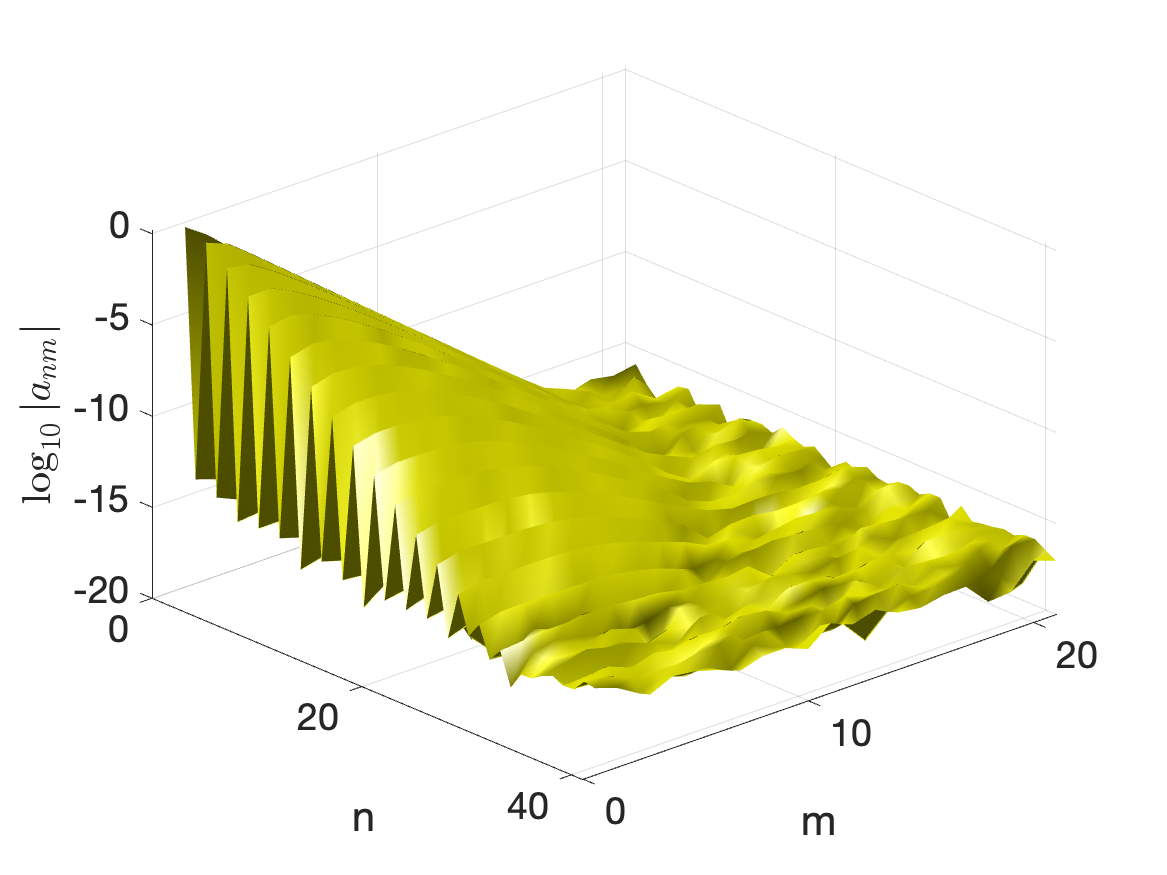}
 \caption{The function $W$ (\ref{rho}) for the double 
 Schwarzschild solution for $m_{1}=m_{2}=1$ and $R_{0}=5$ on the left 
 and the spectral coefficients on the right.}
 \label{figW}
\end{figure}

We introduce the standard Chebyshev collocation points for 
$x\in [-1,1]$ and $\eta\in\eta_{0}[-1,1]$. For the example $R_{0}=5$ and 
$m_{1}=m_{2}=1$ which corresponds to $\mu\sim0.1837$ and $\eta_{0}\sim 2.2597$, 
we show the function $W$ in Fig.~\ref{figW}, where we have normalized 
$W$ to 1 at infinity. With $N_{x}=20$ and $N_{\eta}=40$, the spectral 
coefficients shown on the right of the same figure decrease to machine 
precision.

\begin{figure}[h!]
  \includegraphics[width=0.49\textwidth]{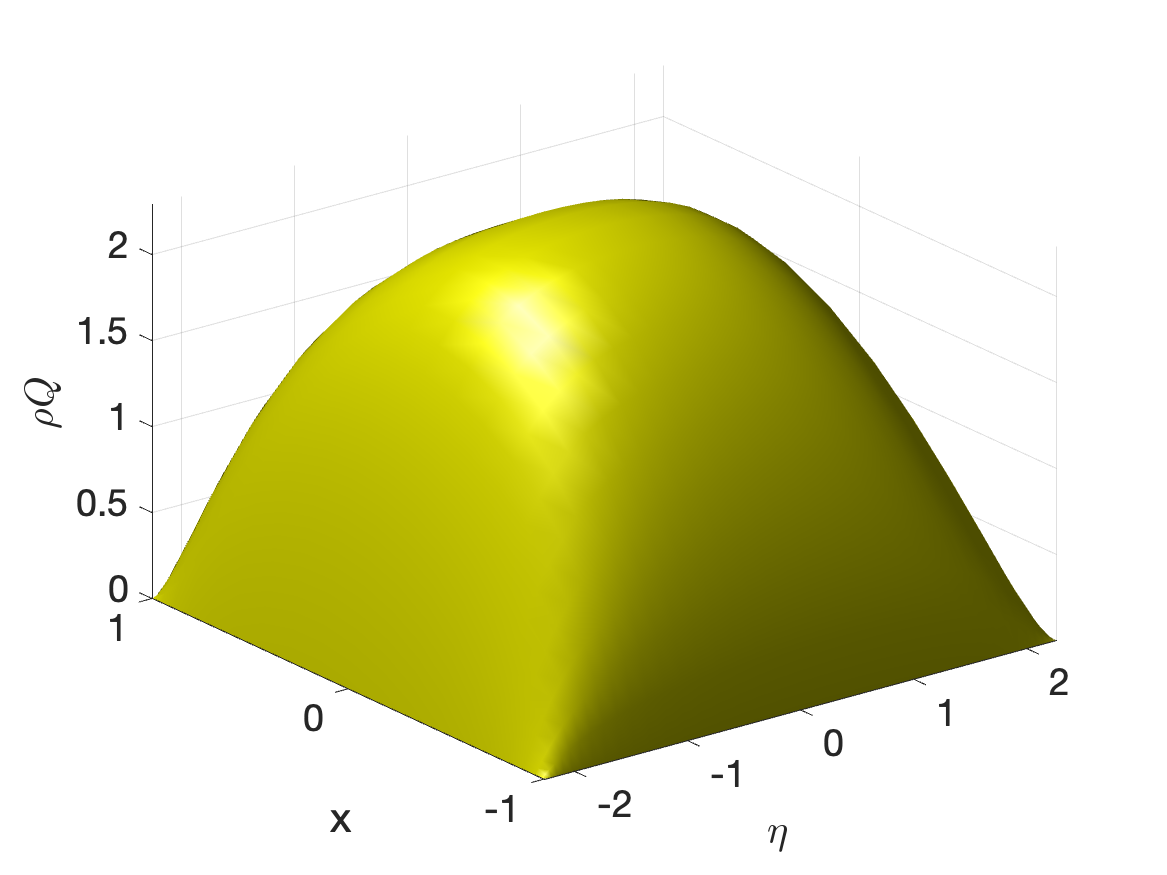}
  \includegraphics[width=0.49\textwidth]{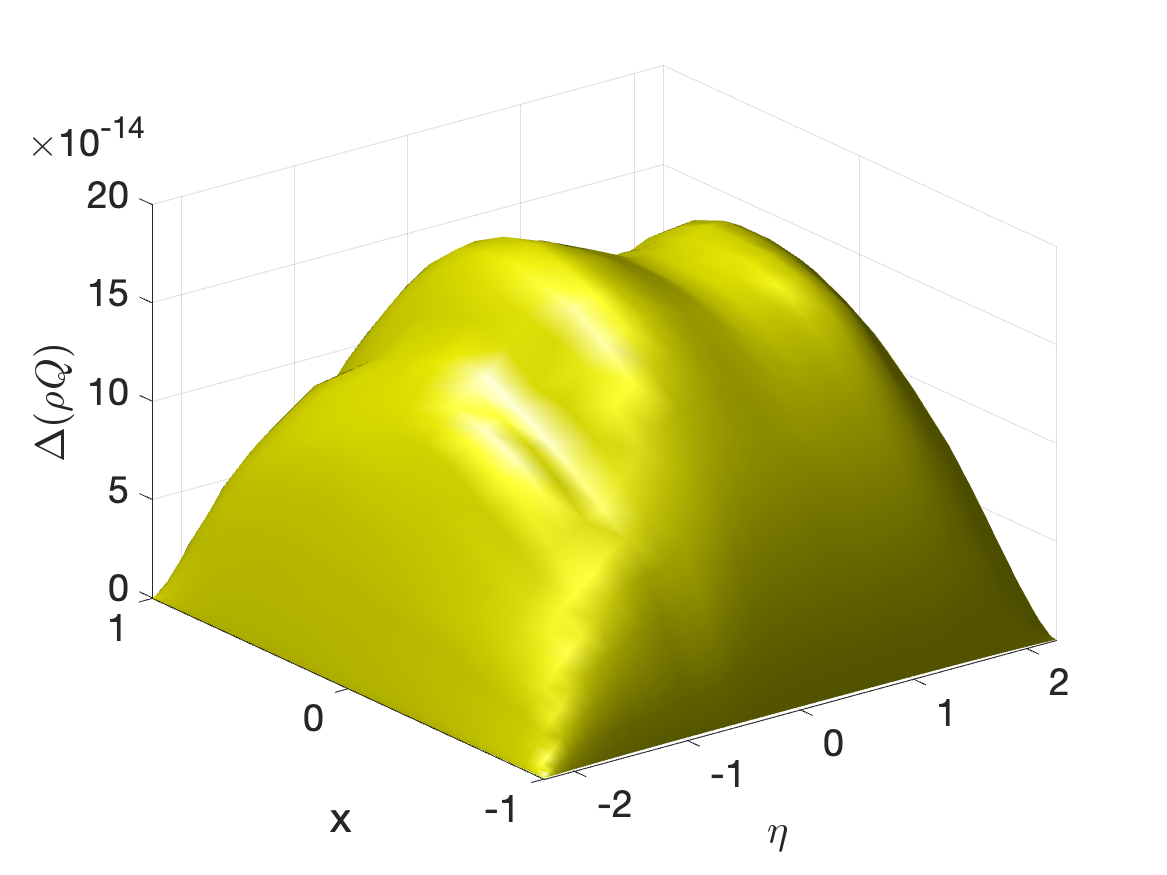}
  \caption{The function $\rho Q$ (\ref{rho}) for the double 
 Schwarzschild solution for $m_{1}=m_{2}=1$ and $R_{0}=5$ on the left 
 and the difference between numerical and exact solution on the right.}
 \label{figrho}
\end{figure}

The corresponding solution for $\rho Q$ is shown on the left of 
Fig.~\ref{figrho}. It vanishes as expected on the boundary of the 
computational domain. The difference between the numerical and the 
exact solution is shown on the right. As expected from the spectral 
coefficients, it is of the order of $10^{-13}$. 

\section{Multi-domain spectral approach}
In this section we will summarise basic concepts of the multi-domain 
spectral approach which is also the basis of Kadath 
\cite{Grandclement2010}. 

\subsection{Domains}
Spectral methods as discussed in the previous section are very 
efficient in approximating smooth functions. In contrast they are 
only of first order in application to discontinuous functions as the 
function $e^{2k}$, where a Gibbs phenomenon will be observed. Since 
this is due to the singular Weyl strut that keeps the spacetime 
static, we will not discuss this case here. 

A possible way to address non-smoothness with a spectral method is to 
introduce several domains where the considered function is smooth in 
each of them. This has also the benefit to reduce the number of 
Chebyshev polynomials needed to numerically resolve a given function 
which is interesting because of the mentioned conditioning problems 
of Chebyshev differentiation matrices. The problem to be addressed 
here is the function $f$ that has a cusp at infinity as can be seen 
in Fig.~\ref{fkbispfig}. This is due to the asymptotic behavior of 
$f\sim 1-2M/r$, where $r=\sqrt{\rho^{2}+z^{2}}$ since $r$ is not a 
smooth function of $\eta$ and $\theta$ near the origin in this 
coordinate system, see (\ref{inf}).

\begin{figure}[h!]
  \includegraphics[width=0.6\textwidth]{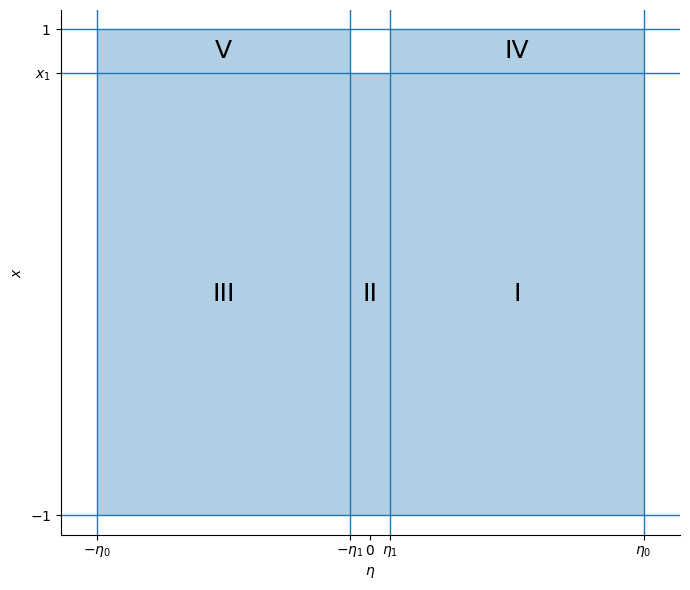}
 \caption{The five numerical domains.}
 \label{kadath_domains}
\end{figure}

To avoid this problem, we cut out a rectangle near infinity similar 
to the approach in \cite{Grandclement2010}, where a curvilinear 
rectangle was used. We introduce in the following positive 
$\eta_{1}\sim 0$ and $x_{1}\sim 1$ and with this the five domains, 
see Fig.~\ref{kadath_domains}:\\
I. $\eta_{1}\leq \eta\leq \eta_{0}$, $-1\leq x\leq x_{1}$;\\
II. $-\eta_{1}< \eta< \eta_{1}$, $-1\leq x\leq x_{1}$;\\
III. $-\eta_{0}\leq \eta\leq -\eta_{1}$, $-1\leq x\leq x_{1}$;\\
IV. $\eta_{1}\leq \eta\leq \eta_{0}$, $x_{1}<x\leq 1$;\\
V. $-\eta_{0}\leq \eta\leq -\eta_{1}$, $x_{1}<x\leq 1$.

At the boundaries between these domains, the functions have to be 
$C^{1}$ in the respective variable, say between domains I and II, the 
functions have to be differentiable in $\eta$. At the 
boundaries to the domain containing infinity, the exact solution is 
imposed. These conditions will 
be again implemented with a $\tau$-method. At the rectangle cut out 
near infinity, the exact solution will be imposed.

\subsection{The Ernst potential $f$}
In order to take 
care of the horizons where $f$ vanishes, we make the ansatz
\begin{equation}
	f = (1-\eta^{2}/\eta_{0}^{2})^{2}e^{U}
	\label{ernst2}
\end{equation}
and get for the Ernst equation
\begin{equation}
	U_{\eta\eta}+(\ln \rho)_{\eta}U_{\eta}-\frac{4}{\eta_{0}^{2}-\eta^{2}}
	-\frac{8\eta^{2}}{(\eta_{0}^{2}-\eta^{2})^{2}}-
	\frac{4\eta}{\eta_{0}^{2}-\eta^{2}}(\ln \rho)_{\eta} 
	+U_{\theta\theta}+(\ln \rho)_{\theta}U_{\theta}=0
	\label{ernst3}.
\end{equation}
This leads with (\ref{rho}) to
\begin{equation}
	\begin{split}
	&U_{\eta\eta}+\left(-\frac{\sinh\eta}{Q}-\frac{2\eta}{\eta_{0}^{2}-\eta^{2}}
	+\frac{W_{\eta}}{W}\right)U_{\eta} + (1-x^{2})U_{xx} \\ & +\left(-2x+\frac{1-x^{2}}{Q}+
	\frac{(1-x^{2})W_{x}}{W}\right)U_{x} = \frac{4}{\eta_{0}^{2}-\eta^{2}}\left(
	1-\frac{\eta\sinh\eta}{Q}+\frac{\eta W_{\eta}}{W}\right),
	\end{split}
\label{ernst4}
\end{equation}
a singular linear equation with a source term which will be solved 
for $U$ with a vanishing condition at infinity  
(U vanishes at infinity since f goes to 1 there). But since we cut 
out the white rectangle in Fig.~\ref{kadath_domains} containing 
infinity, we impose instead boundary conditions at the common 
boundaries of the rectangle with domains II, IV and V. We use the 
exact solution for U as a boundary condition. 

\begin{figure}[h!]
  \includegraphics[width=0.32\textwidth]{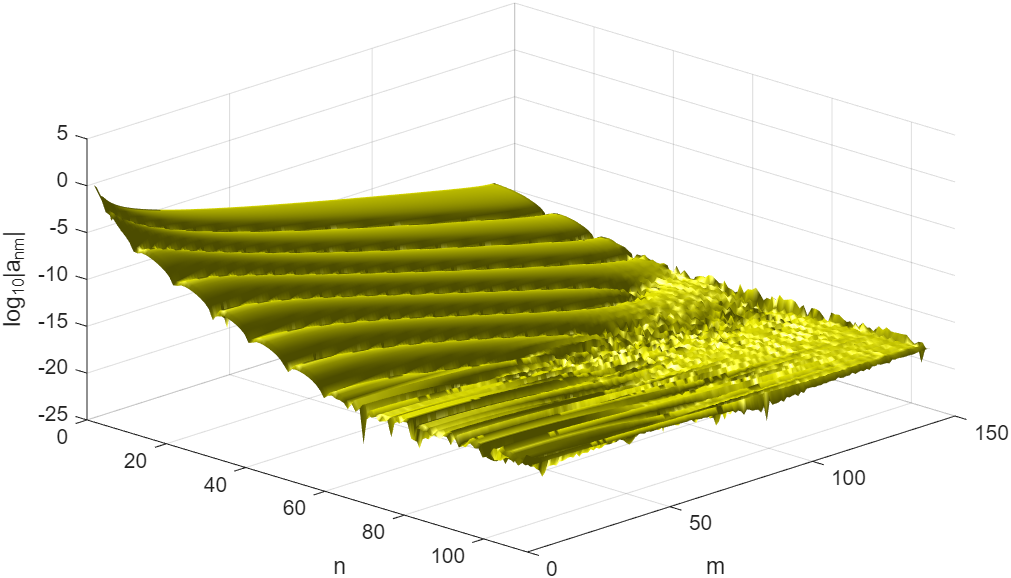}
  \includegraphics[width=0.32\textwidth]{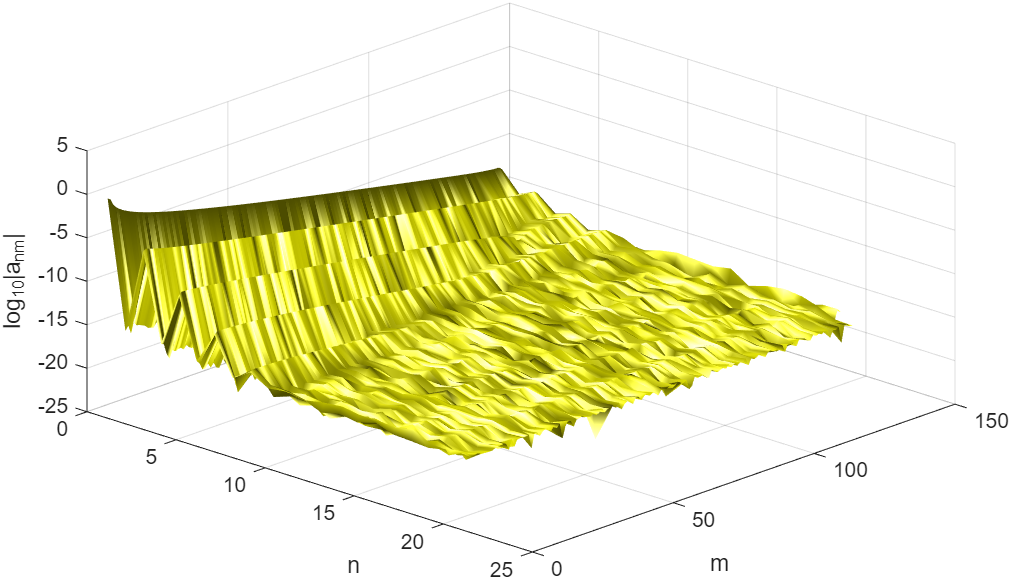}
  \includegraphics[width=0.32\textwidth]{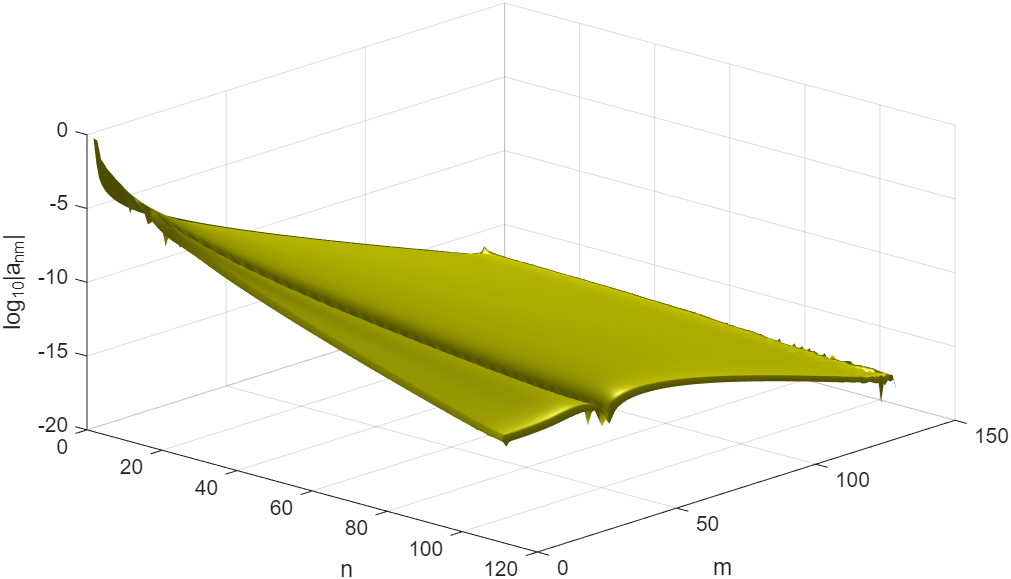}
  \caption{The spectral coefficients for $U$ in the domains I (left), 
  II (middle) and IV (right), for $m_{1}=m_{2}=1$ and $R_{0}=5$.}
 \label{Ucoef}
\end{figure}

In Fig.~\ref{Ucoef} we show the spectral coefficients for the exact 
solution $U$ in domains I, II and IV (for symmetry reasons we do not show 
the ones in domain III and IV). In all domains we apply $150$ Chebyshev 
polynomials in $x$, and in  domain I $100$ in $\eta$, in domain II  
$20$ in $\eta$, and in domain IV  100 in $\eta$. This gives an indication 
of the needed spectral resolution in the studied example. 

\subsection{Numerical solution for $U$}
\begin{figure}[h!]
  \includegraphics[width=0.49\textwidth]{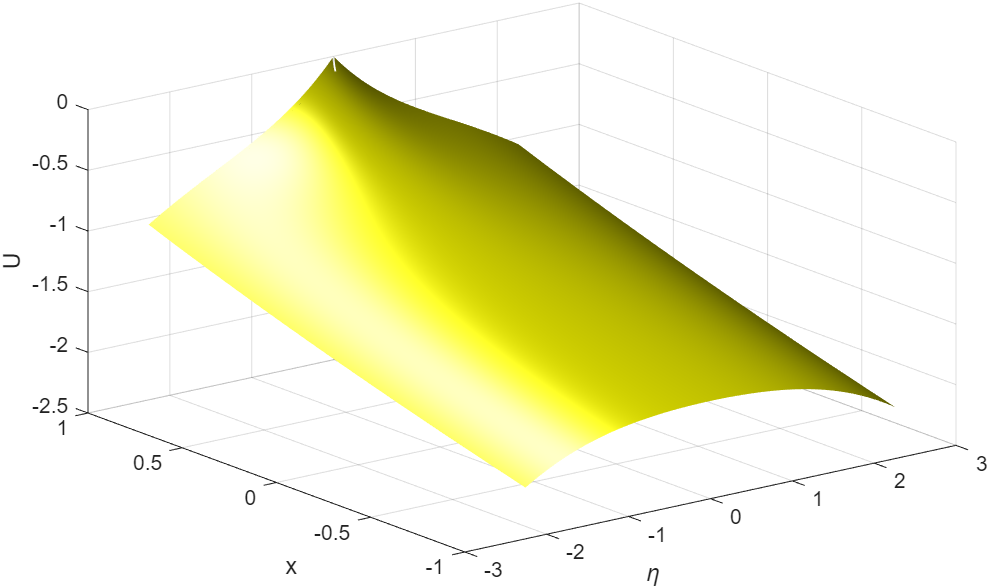}
  \includegraphics[width=0.49\textwidth]{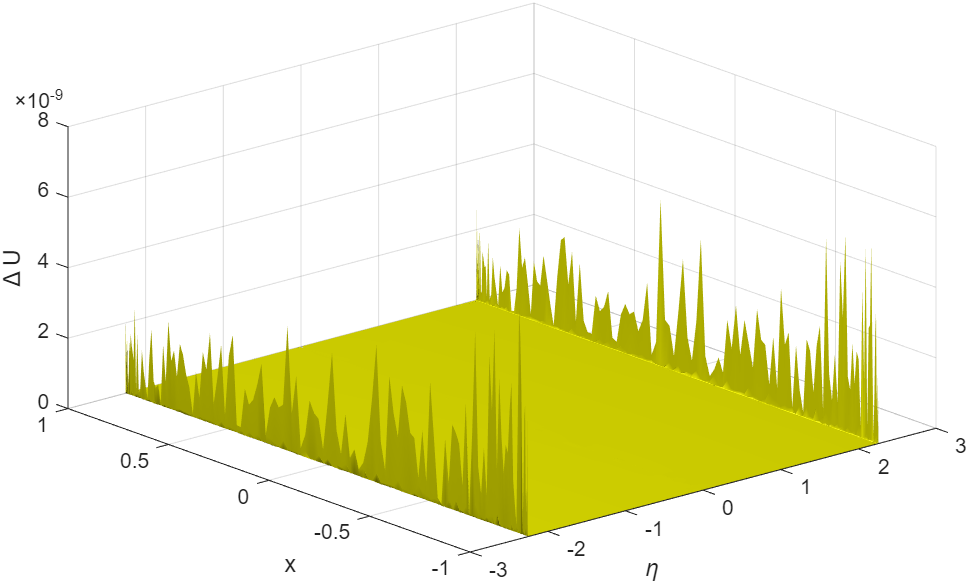}
  \caption{The function $U$ (\ref{ernst2}) for the double 
 Schwarzschild solution for $m_{1}=m_{2}=1$ and $R_{0}=5$ on the left, 
 and the difference between numerical and exact solution on the right.}
 \label{figU}
\end{figure}

The numerical solution with the above numerical parameters can be 
seen in Fig.~\ref{figU} on the left. The small cut out domain near 
infinity is visible. The more one avoids the cusp singularity at 
infinity, i.e., the larger $\eta_{1}$ and the smaller $x_{1}$ are 
chosen,  the lower the needed resolution. The difference between the 
exact and the numerical solution is shown on the right of 
Fig.~\ref{figU}. For the shown parameters, it is of the order of 
$10^{-9}$, the main difference appearing at the horizon where the 
Ernst potential vanishes. 

\begin{figure}[h!]
  \includegraphics[width=0.49\textwidth]{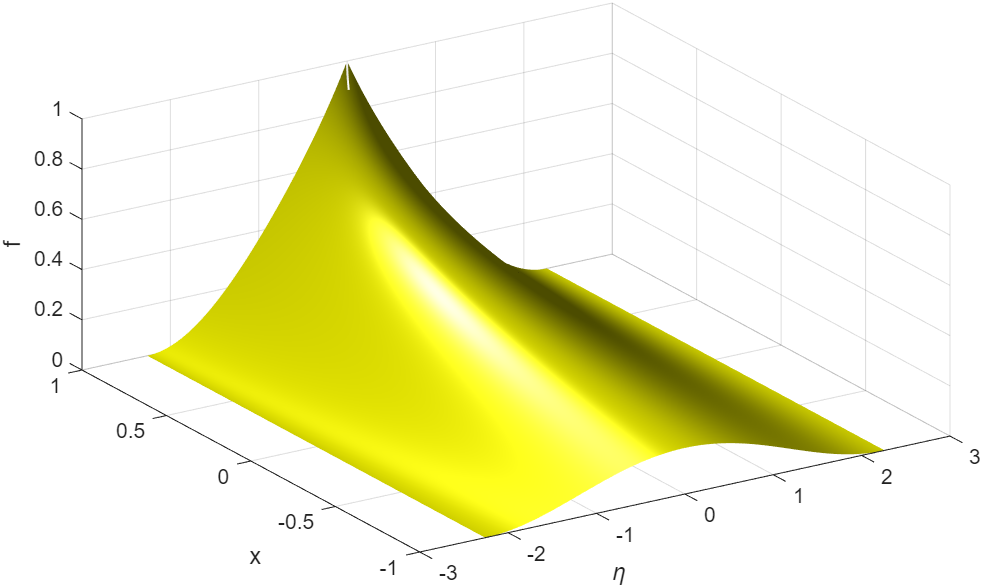}
  \includegraphics[width=0.49\textwidth]{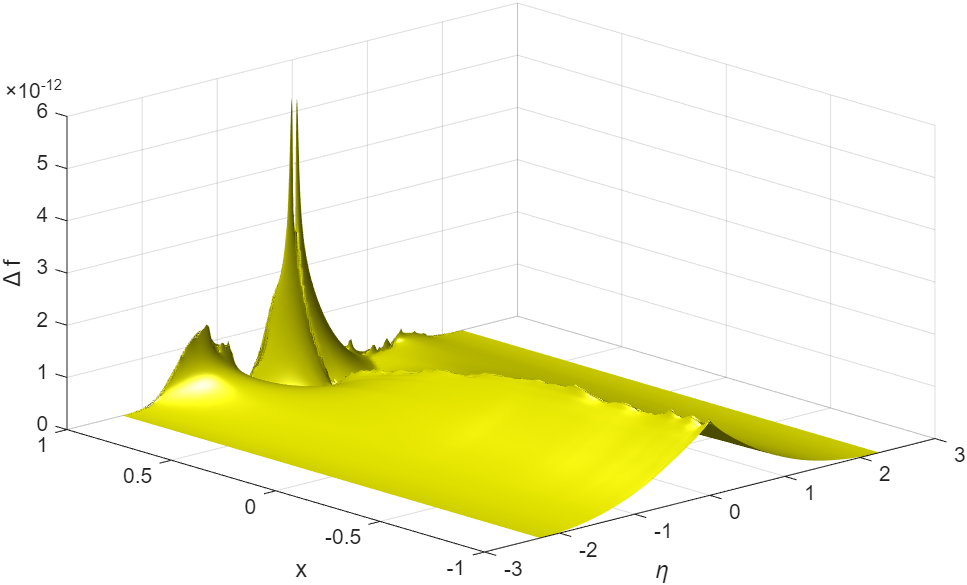}
  \caption{The function $f$ (\ref{ernst2}) for the double 
 Schwarzschild solution for $m_{1}=m_{2}=1$ and $R_{0}=5$ on the left, 
 and the difference between numerical and exact solution on the right.}
 \label{figf}
\end{figure}

The solution for the Ernst equation itself is shown in Fig.~\ref{figf} on the 
left. The difference between the exact and the numerical solution is 
of the order of $10^{-12}$ as can be seen on the right of the same 
figure. Here the main error is near the infinite domain. 

\section{Outlook}
In this paper we have given an explicit representation of the double 
Schwarzschild solution in bispherical coordinates in terms of Jacobi 
elliptic functions. The boundary of the computational domain is given 
by the horizons and the symmetry axis, infinity corresponding to a point of the 
boundary. The Einstein equations reduce in this case to 
$h_{\phi\phi}$ being given by a harmonic function with a vanishing 
condition on the boundary of the computational domain and the 
Euler-Darboux equation for the logarithm of the (real) Ernst 
potential. If the behavior of these two functions at the horizons and 
the axis is addressed by an explicit ansatz for these functions, the 
Einstein equations reduce to a set of coupled linear equations with 
Fuchsian singularities at the horizons and the axis. 

These two equations were solved with a multi-domain spectral method 
which led to a reproduction of the exact solution to the order of 
$10^{-12}$. 
The third metric 
function is given in terms of 
quadratures of the first two functions. Since it is discontinuous due to the singular Weyl strut 
separating the two horizons, it was not considered here. 

This work is intended to be a preparatory step to treat binary 
black holes with a helical Killing vector in the Ernst formulation, 
see \cite{klein2004}. The idea is to make a similar ansatz for the 
Ernst potential and the metric $h$ near the horizon which will lead 
to a Fuchsian nonlinear system. This will be solved as in 
\cite{Bai2016} with a modified Newton iteration to take care of the 
light cylinder where the equations are also singular. Near infinity, 
a rectangle will be cut out as in \cite{Grandclement2010} and the 
present paper where we impose either a type N solution obtained by solving the 
linearized Einstein equations in the presence of a helical  Killing 
vector as in \cite{klein2004} or a standard asymptotically flat 
metric (this would assume that the helical symmetry only holds in the 
vicinity of the black holes). This will be the subject of future 
work. 

\appendix

\section{Construction of the conformal map}

In the equal-mass double Schwarzschild case, it is possible to construct a conformal map $w$ such that
\[
\rho + \ii z = w(\eta+\ii\theta),
\]
where $(\rho,z)$ ranges over
\[
\mathcal{D}_w := \mathbb{R}^+\times\mathbb{R}
\]
as $(\eta,\theta)$ varies in
\[
\mathcal{D}_b := [-\eta_0,\eta_0]\times[0,\pi]\setminus\{0\},
\]
for some parameter $\eta_0>0$. Since Weyl coordinates are unbounded, the map $w$ must have a pole, representing spatial infinity in bispherical coordinates. By axial symmetry, one may identify $\psi$ with $\phi$. Moreover, these coordinates should represent the two horizons as the spheres $\{\eta=\eta_0\}$ and $\{\eta=-\eta_0\}$.

The problem is therefore to construct a conformal map from $\mathcal{D}_b$ onto $\mathcal{D}_w$, that is, a map from a rectangle to the right half-plane. In order for the horizons to be represented by two spheres, the two vertical edges of $\mathcal{D}_b$ must be mapped to two segments of the imaginary axis. The remaining edges must be mapped to the complementary parts of the imaginary axis. This requires
\[
\Re(w(u))=0,
\qquad\text{for all }u\in\partial\mathcal{D}_b\setminus\{0\},
\]
where
\[
u:=\eta+\ii\theta.
\]

A convenient way to proceed is to use the analytic continuation of the Jacobi elliptic function $\sn$, see \cite{Lawden}. For all $v\in\mathbb{R}$, define
\[
    \sn(v,\mu) := \sin\bigl(A(v,\mu)\bigr),
\]
where
\[
A(v,\mu) := \int_0^v \frac{dt}{\sqrt{1-\mu\sin^2 t}}
\]
is the incomplete elliptic integral of the first kind, and $\mu\in(0,1)$ is the modulus. Let
\[
K(\mu):=A(\pi/2,\mu)
\]
denote the complete elliptic integral of the first kind, and define
\[
K'(\mu):=K(1-\mu).
\]
It can then be shown that $\sn(\cdot,\mu)$ conformally maps the domain
\[
\mathcal{D}:=[-K,K]\times[-K',0]\setminus\{-\ii K'\}
\]
onto the lower half-plane, with a pole at $-\ii K'$. Moreover, the boundary $\partial\mathcal{D}$ is mapped as follows:
\[
    \begin{array}{llll}
        \sn\bigl([-K,K]\times\{0\},\mu\bigr) = [-1,1]\times\{0\}, \\[1.2ex]
        \sn\bigl(\{K\}\times[-K',0],\mu\bigr) = [1,1/\sqrt{\mu}]\times\{0\}, \\[1.2ex]
        \sn\bigl([-K,K]\times\{-K'\},\mu\bigr) = \bigl(]-\infty,-1/\sqrt{\mu}]\cup[1/\sqrt{\mu},+\infty[\bigr)\times\{0\}, \\[1.2ex]
        \sn\bigl(\{-K\}\times[-K',0],\mu\bigr) = [-1/\sqrt{\mu},-1]\times\{0\}.
    \end{array}
\]

To obtain the right half-plane, it is sufficient to multiply by $\ii$, which corresponds to a rotation of the image by $\pi/2$. Since the map $w$ is defined on $\mathcal{D}_b$, one first introduces a linear transformation
\[
v=\alpha u+\beta
\]
mapping $\mathcal{D}_b$ onto $\mathcal{D}$. A direct computation shows that
\[
    \alpha = \frac{K}{\eta_0} = \frac{K'}{\pi},
    \qquad
    \beta = -iK'.
\]
Furthermore, since
\[
z(\eta_0)=\frac{R_0}{2}+m
\qquad\text{and}\qquad
z(\eta_0 + \ii\pi)=\frac{R_0}{2}-m,
\]
the map must be scaled by the factor $\dfrac{R_0}{2}-m$. This leads to the relation
\[
    \frac{R_0/2-m}{\sqrt{\mu}} = \frac{R_0}{2}+m,
\]
from which one obtains
\[
\mu = \left(\frac{R_0-2m}{R_0+2m}\right)^2,
\qquad
\eta_0 = \frac{\pi K(\mu)}{K'(\mu)}.
\]
Hence the conformal map is given by
\[
    w(u) = \ii\left(\frac{R_0}{2}-m\right)\sn\left(\frac{K}{\eta_0}u - \ii K', \mu\right).
\]
Using the identity
\[
\sn(v- \ii K'(\mu),\mu) = \bigl(\sqrt{\mu}\,\sn(v,\mu)\bigr)^{-1},
\]
this expression can be rewritten in the simpler form
\begin{equation}
    w(u) = \ii\left(\frac{R_0}{2}+m\right)\ns\left(\frac{K}{\eta_0}u, \mu\right),
    \label{w(u)app}
\end{equation}
where $\ns := 1/\sn$. Finally, we define
\[
\rho(\eta,\theta) := \Re\bigl(w(\eta+\ii\theta)\bigr),
\qquad
z(\eta,\theta) := \Im\bigl(w(\eta+\ii\theta)\bigr).
\]
The map $w$ has a pole at $u=0$, which corresponds to spatial infinity.

\begin{figure}[H]
\includegraphics[width=0.98\textwidth]{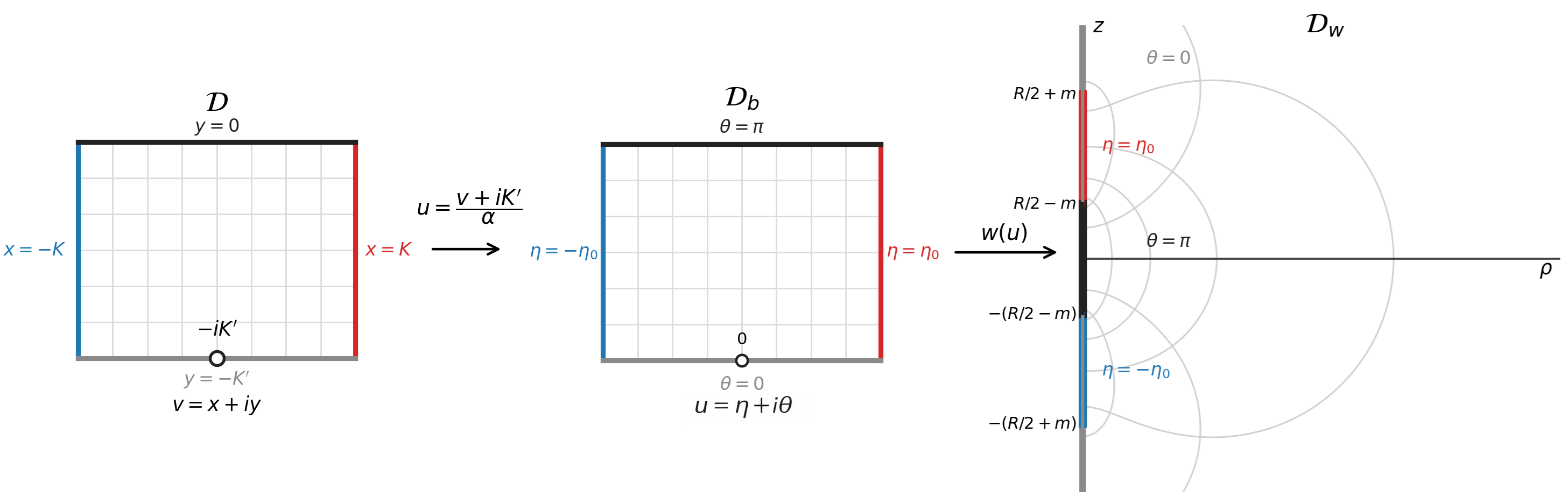}
 \caption{Mapping from the (half) fundamental domain $\mathcal{D}$ to the Weyl coordinate domain $\mathcal{D}_w$ through the bispherical domain $\mathcal{D}_b$.}
 \label{conformal_map}
\end{figure}

Fig.~\ref{conformal_map} above shows how horizontal and vertical lines in the domain $\mathcal{D}_b$ are transformed by the conformal map $w$. However, this map can also be represented using a domain-coloring plot, as shown below in Fig.~\ref{w_plot}. The colors represent the phase, and the white lines represent curves of equal amplitude. One can clearly see the pole at $u = 0$ and the zero at $u = i\pi$.

\begin{figure}[H]
\includegraphics[width=0.6\textwidth]{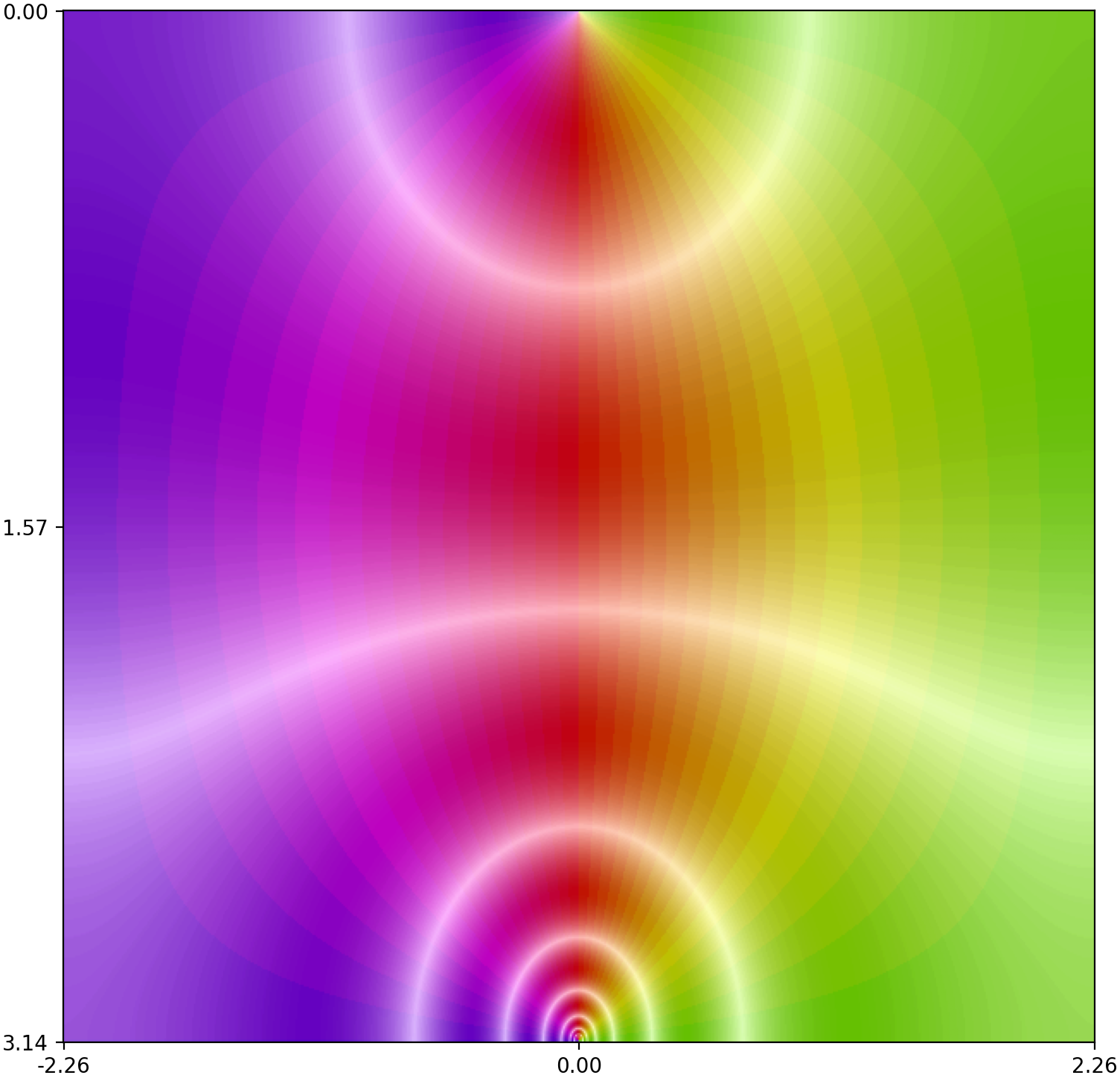}
\caption{Domain-coloring plot of the conformal map $w$ in the domain $\mathcal{D}_b$. The $\theta$ axis is reversed in this figure.}
\label{w_plot}
\end{figure}

\section{Uniqueness of the conformal map}

The map $w$ is in fact the unique conformal map that preserves the structure of the horizons in Weyl coordinates. Indeed, suppose that there exists another conformal map
\[
    \tilde{w}:\mathcal{D}_b\to\mathcal{D}_w
\]
such that
\[
    \begin{array}{ll}
        \tilde{w}(\eta_0) = \ii\left(\frac{R_0}{2}+m\right), & \tilde{w}(\eta_0 + \ii\pi) = \ii\left(\frac{R_0}{2}-m\right),
        \\[1.2ex]
        \tilde{w}(-\eta_0) = -\ii\left(\frac{R_0}{2}+m\right), & \tilde{w}(-\eta_0 + \ii\pi) = -\ii\left(\frac{R_0}{2}-m\right).
    \end{array}
\]
Then the map
\[
    \tilde{w}\circ w^{-1}:\mathcal{D}_w\to\mathcal{D}_w
\]
is a conformal automorphism of the half-plane fixing the four points
\[
    \left\{
    \ii\left(\frac{R_0}{2}+m\right),\,
    \ii\left(\frac{R_0}{2}-m\right),\,
    -\ii\left(\frac{R_0}{2}+m\right),\,
    -\ii\left(\frac{R_0}{2}-m\right)
    \right\}.
\]
Since every conformal automorphism of the half-plane is a M\"obius transformation, and since $\tilde{w}\circ w^{-1}$ fixes at least three distinct points (for $m>0$), it follows that
\[
    \tilde{w}\circ w^{-1} = \mathrm{id}_{\mathcal{D}_w}.
\]
Therefore, $\tilde{w}=w$, which proves the uniqueness of the conformal map.

\section{Behavior near infinity}

In bispherical coordinates, spatial infinity corresponds to the pole 
$u=0$, where $r := \sqrt{\rho^2 + z^2}$ tends to infinity. In order 
to study the behavior of the coordinates and the metric near infinity, one uses the expansion
\[
    \sn\left(\alpha u,\mu\right) \underset{u\to 0}{=} \alpha u + \mathcal{O}(|u|^3).
\]
Therefore, at leading order, one has
\[
    \rho(\eta,\theta)\underset{u\to 0}{=}\frac{R_0/2 + m}{\alpha}\frac{\theta}{\eta^2 + \theta^2} + \mathcal{O}(1),
    \qquad
    z(\eta,\theta) \underset{u\to 0}{=} \frac{R_0/2 + m}{\alpha}\frac{\eta}{\eta^2 + \theta^2} + \mathcal{O}(1).
\]
From \eqref{cart_bis}, one sees that by choosing
\[
    a = \frac{R_0/2 + m}{2\alpha},
\]
one obtains a natural interpretation of these bispherical coordinates: sufficiently far from the black holes, they reduce to the standard bispherical coordinate transformation in flat space, since $f,\;e^{2k}\to 1$ and, from \eqref{dw2},
\[
    h_{\eta\eta} = h_{\theta\theta} = |w'(\eta+\ii\theta)|^2e^{2k} \underset{u\to 0}{=} \frac{4a^2}{(\eta^2+\theta^2)^2} + \mathcal{O}(|u|^{-2}) \underset{u\to 0}{=} \frac{a^2}{Q^2} + \mathcal{O}(|u|^{-2}).
\]
Here $h_{ab}$ is the spatial metric defined in \eqref{space_metric}. The azimuthal component can be expressed as
\[
    h_{\phi\phi} = \rho^2 \underset{u\to 0}{=} \dfrac{4a^2\theta^2}{(\eta^2 + \theta^2)^2} + \mathcal{O}(|u|^{-2}) \underset{u\to 0}{=} \dfrac{a^2\sin^2\theta}{Q^2} + \mathcal{O}(|u|^{-2}).
\]
One thus recovers the flat metric in standard bispherical coordinates, i.e.\ the metric is asymptotically flat. This justifies why it is convenient to normalize the Weyl coordinates by the factor $Q$ in Figure \ref{norm_weyl}.

\section{Behavior on the axis}

In bispherical coordinates, the axis is given by $\theta = 0$ in the exterior region and by $\theta = \pi$ between the horizons. Thus, to study the behavior of the metric near the axis, one has to examine how the conformal map $w$ in \eqref{w(u)app} behaves at $\theta = 0$ and $\theta = \pi$. First, let us consider the following expansion:
\[
    \text{ns}(\alpha u, \mu) \underset{\theta\to 0^+}{=} \text{ns}(\alpha\eta, \mu) - i\alpha\text{cn}(\alpha\eta, \mu)\text{dn}(\alpha\eta, \mu)\text{ns}^2(\alpha\eta, \mu)\theta + \mathcal{O}(\theta^2).
\]
Therefore, $\rho(u) = \Re(w(u))$ and $z(u) = \Im(w(u))$ can be expressed at leading order as
\[
    \begin{array}{cc}
       \rho(u) \underset{\theta\to 0^+}{=} 2a\alpha^2\text{cn}(\alpha\eta, \mu)\text{dn}(\alpha\eta, \mu)\text{ns}^2(\alpha\eta, \mu)\theta + \mathcal{O}(\theta^2), \qquad &
       z(u) \underset{\theta\to 0^+}{=} 2a\alpha\text{ns}(\alpha\eta, \mu) + \mathcal{O}(\theta^2).
    \end{array}
\]
Since $w$ is conformal, one can write its derivative as $w'(u) = z_{\theta} - \ii\rho_{\theta}$. Therefore, to leading order,
\[
    w'(u) \underset{\theta\to 0^+}{=} -2\ii a\alpha^2\text{cn}(\alpha\eta, \mu)\text{dn}(\alpha\eta, \mu)\text{ns}^2(\alpha\eta, \mu) + \mathcal{O}(\theta).
\]
It is well known that, on the axis, the Weyl potential $e^{2k}$ is constant along the axis and is equal to $1$ in the exterior region. Hence the metric takes the form
\[
    \begin{array}{ll}
        h_{\eta\eta} = h_{\theta\theta} = |w'(\eta + \ii\theta)|^2e^{2k} \underset{\theta\to 0^+}{=} \Omega_0^2(\eta) + \mathcal{O}(\theta), \\\\
        h_{\phi\phi} = \rho^2 \underset{\theta\to 0^+}{=} \Omega_0^2(\eta)\theta^2 + \mathcal{O}(\theta^3) \underset{\theta\to 0^+}{=} \Omega_0^2(\eta)\sin^2\theta + \mathcal{O}(\theta^3),
    \end{array}
\]
with $\Omega_0^2(\eta) := 4a^2\alpha^4\text{cn}^2(\alpha\eta, 
\mu)\text{dn}^2(\alpha\eta, \mu)\text{ns}^4(\alpha\eta, \mu)$. One 
can see that the metric is locally elementary flat near the axis, with $\Omega_0$ as the conformal factor. Next, let us consider the axis between the horizons. The expansion of $\text{ns}(\alpha u, \mu)$ at $\theta = \pi$ is
\[
    \text{ns}(\alpha u, \mu) \underset{\theta\to \pi^-}{=} \text{ns}(\alpha\eta + \ii K', \mu) - i\alpha\text{cn}(\alpha\eta + \ii K', \mu)\text{dn}(\alpha\eta + \ii K', \mu)\text{ns}^2(\alpha\eta + \ii K', \mu)(\theta - \pi) + \mathcal{O}\left(|\theta - \pi|^2\right),
\]
where the relation $\dfrac{K}{\eta_0} = \dfrac{K'}{\pi}$ has been used. One can simplify this further by using the identities, valid for all $x \in \mathbb{R}$, see \cite{Lawden},
\[
    \begin{array}{ll}
        \text{ns}(x + \ii K', \mu) = \sqrt{\mu}\text{sn}(x, \mu), &
        \text{cn}(x + \ii K', \mu) = -\dfrac{\ii}{\sqrt{\mu}}\text{ns}(x, \mu)\text{dn}(x, \mu), \\
        \text{dn}(x + \ii K', \mu) = -\ii\text{ns}(x, \mu)\text{cn}(x, \mu),
    \end{array}
\]
and obtain
\[
    \text{ns}(\alpha u, \mu) \underset{\theta\to \pi^-}{=} \sqrt{\mu}\text{sn}(\alpha\eta, \mu) + \ii\alpha\sqrt{\mu}\text{cn}(\alpha\eta, \mu)\text{dn}(\alpha\eta, \mu)(\theta - \pi) + \mathcal{O}\left(|\theta - \pi|^2\right).
\]
Therefore,
\[
    \begin{array}{ll}
       \rho(u) \underset{\theta\to \pi^-}{=} 2a\alpha^2\sqrt{\mu}\text{cn}(\alpha\eta, \mu)\text{dn}(\alpha\eta, \mu)(\pi - \theta) + \mathcal{O}\left(|\theta - \pi|^2\right), \\\\
       z(u) \underset{\theta\to \pi^-}{=} 2a\alpha\sqrt{\mu}\text{sn}(\alpha\eta, \mu) + \mathcal{O}\left(|\theta - \pi|^2\right).
    \end{array}
\]
In the same way, one finds the expression for $w'(u)$ near the axis:
\[
    w'(u) \underset{\theta\to \pi^-}{=} 2ia\alpha^2\sqrt{\mu}\text{cn}(\alpha\eta, \mu)\text{dn}(\alpha\eta, \mu) + \mathcal{O}\left(|\theta - \pi|\right).
\]
It is possible to compute the potential $e^{2k}$ on the axis between the horizons using \eqref{fk} for $\rho = 0$ and $z \in [-R_0/2 + m, R_0/2 - m]$, and one finds that
\[
e^{2k} = \left(1 - \dfrac{4m^2}{R_0^2}\right)^2.
\]
One therefore obtains the metric components
\[
    \begin{array}{ll}
        h_{\eta\eta} = h_{\theta\theta} = |w'(\eta + \ii\theta)|^2e^{2k} \underset{\theta\to \pi^-}{=} \left(1 - \dfrac{4m^2}{R_0^2}\right)^2\Omega_1^2(\eta) + \mathcal{O}(|\theta - \pi|), \\\\
        h_{\phi\phi} = \rho^2 \underset{\theta\to \pi^-}{=} \Omega_1^2(\eta)(\pi - \theta)^2 + \mathcal{O}(|\theta - \pi|^3) \underset{\theta\to \pi^-}{=} \Omega_1^2(\eta)\sin^2\theta + \mathcal{O}(|\theta - \pi|^3),
    \end{array}
\]
with $\Omega_1^2(\eta) := 4a^2\alpha^4\mu\text{cn}^2(\alpha\eta, \mu)\text{dn}^2(\alpha\eta, \mu)$. Since $\left(1 - \dfrac{4m^2}{R_0^2}\right)^2 < 1$, the metric is no longer locally conformally flat near the axis between the horizons; this naturally corresponds to the Weyl strut.

\section{Behavior on the horizons}

In bispherical coordinates, the horizons are defined by $|\eta| = \eta_0$. The following discussion focuses only on the case $\eta = \eta_0$; the other case is symmetric. Let $\delta := \eta - \eta_0$. The expansion of $\text{ns}(\alpha u, \mu)$ at $\delta = 0$ will be used:
\begin{align*}
    \text{ns}(\alpha u, \mu) = \text{ns}(K + \ii\alpha\theta + \alpha\delta, \mu) \underset{\delta\to 0^-}{=} \text{ns}(K + \ii\alpha\theta, \mu) - \alpha\text{cn}(K + \ii\alpha\theta, \mu)\text{dn}(K + \ii\alpha\theta, \mu)\text{ns}^2(K + \ii\alpha\theta, \mu)\delta \\
    + \dfrac{1}{2}\alpha^2\text{ns}(K + \ii\alpha\theta, \mu)\left(2\text{ns}^2(K + \ii\alpha\theta, \mu) - 1 - \mu\right)\delta^2 + \mathcal{O}(\delta^3),
\end{align*}
where $\alpha = \dfrac{K}{\eta_0}$ has been used in the first equality. Then, by using the quarter-period shift and the pure imaginary argument formulas, see \cite{Lawden}, one can show that
\[
    \begin{array}{ccc}
       \text{ns}(K + \ii\alpha\theta, \mu) = \text{dn}(\alpha\theta, \mu'), &
       \text{cn}(K + \ii\alpha\theta, \mu) = -\ii\sqrt{\mu'}\dfrac{\text{sn}(\alpha\theta, \mu')}{\text{dn}(\alpha\theta, \mu')}, &
       \text{dn}(K + \ii\alpha\theta, \mu) = \sqrt{\mu'}\dfrac{\text{cn}(\alpha\theta, \mu')}{\text{dn}(\alpha\theta, \mu')},
    \end{array}
\]
with $\mu' := 1 - \mu$. Hence
\[
    \text{ns}(\alpha u, \mu) \underset{\delta\to 0^-}{=} \text{dn}(\alpha\theta, \mu') + \ii\alpha\mu'\text{sn}(\alpha\theta, \mu')\text{cn}(\alpha\theta, \mu')\delta + \dfrac{1}{2}\alpha^2\text{dn}(\alpha\theta, \mu')\left(2\text{dn}^2(\alpha\theta, \mu') - 1 - \mu\right)\delta^2 + \mathcal{O}(\delta^3),
\]
and therefore
\[
    \begin{array}{ll}
       \rho(u) \underset{\delta\to 0^-}{=} -2a\alpha^2\mu'\text{sn}(\alpha\theta, \mu')\text{cn}(\alpha\theta, \mu')\delta + \mathcal{O}(\delta^3), \\\\
       z(u) \underset{\delta\to 0^-}{=} 2a\alpha\text{dn}(\alpha\theta, \mu') + a\alpha^3\text{dn}(\alpha\theta, \mu')\left(2\text{dn}^2(\alpha\theta, \mu') - 1 - \mu\right)\delta^2 + \mathcal{O}(\delta^3).
    \end{array}
\]
Since $w$ is conformal, one can write its derivative as $w'(u) = \rho_{\eta} + \ii z_{\eta}$. Therefore, to leading order,
\[
    w'(u) \underset{\delta\to 0^-}{=} -2a\alpha^2\mu'\text{sn}(\alpha\theta, \mu')\text{cn}(\alpha\theta, \mu') + 2\ii a\alpha^3\text{dn}(\alpha\theta, \mu')\left(2\text{dn}^2(\alpha\theta, \mu') - 1 - \mu\right)\delta + \mathcal{O}(\delta^2).
\]
In order to determine how the metric behaves near the horizon, one must examine how the Weyl potential $e^{2k}$ behaves as $\delta \to 0^-$. From the expressions of $A$ and $B$ in \eqref{AB} and the expressions of $\rho$ and $z$ above, one finds
\[
    \begin{array}{ll}
       A - B \underset{\delta\to 0^-}{=} 32a^4\alpha^6\mu'\left(\text{dn}(\alpha\theta, \mu') + \sqrt{\mu}\right)^3\delta^2 + \mathcal{O}(\delta^4), \\\\
       A + B \underset{\delta\to 0^-}{=} 128a^4\alpha^4\mu'\left(\text{dn}(\alpha\theta, \mu') + \sqrt{\mu}\right) + \mathcal{O}(\delta^2).
    \end{array}
\]
From the expressions of $f$ and $e^{2k}$, one obtains the simple expansions near the horizon:
\[
    \begin{array}{ll}
       f \underset{\delta\to 0^-}{=} \dfrac{1}{4}\alpha^2\left(\text{dn}(\alpha\theta, \mu') + \sqrt{\mu}\right)^2\delta^2 + \mathcal{O}(\delta^3), \\\\
       e^{2k} \underset{\delta\to 0^-}{=} \dfrac{16a^4\alpha^6\left(\text{dn}(\alpha\theta, \mu') + \sqrt{\mu}\right)^4}{R_0^4\text{sn}^2(\alpha\theta, \mu')\text{cn}^2(\alpha\theta, \mu')}\delta^2 + \mathcal{O}(\delta^3).
    \end{array}
\]
This gives the following expressions for the metric:
\[
    \begin{array}{ll}
        h_{\eta\eta} = h_{\theta\theta} = |w'(u)|^2e^{2k} \underset{\delta\to 0^-}{=} \dfrac{64a^6\alpha^{10}\mu'^2\left(\text{dn}(\alpha\theta, \mu') + \sqrt{\mu}\right)^4}{R_0^4}\delta^2 + \mathcal{O}(\delta^3), \\\\
        h_{\phi\phi} = \rho^2 \underset{\delta\to 0^-}{=} 4a^2\alpha^4\mu'^2\text{sn}^2(\alpha\theta, \mu')\text{cn}^2(\alpha\theta, \mu')\delta^2 + \mathcal{O}(\delta^4).
    \end{array}
\]
\bibliographystyle{ieeetr}
\bibliography{biblio}
\addcontentsline{toc}{section}{Bibliography}

\end{document}